

\input harvmac
\overfullrule=0pt
\def\pn{\par\noindent}
\def\secskip{\vskip0.9cm}

\def\subskip{\vskip 0.6cm \noindent}
\def\Log{{\rm Log}}
\def\arg{{\rm arg}}
\def\hlf{{1\over 2}}
\def\qtr{{1\over 4}}

\def\Null#1{}

\def\pn{\par\noindent}

\def\subskip{\vskip 0.3cm}
\def\sect#1{\vskip 1cm \centerline{\bf #1} }

\def\Log{{\rm log}}
\def\arg{{\rm arg}}

\def\rPaI{1}
\def\rBax{2}
\def\rBP{3}
\def\rKBP{4}
\def\rKSZ{5}
\def\rKPO{6}
\def\rKPT{7}
\def\rJKM{8}
\def\rBCN{9}
\def\rAf{10}
\def\rKi{11}
\def\rKu{12}
\def\rBRO{13}
\def\rBRT{14}
\def\rAM{15}
\def\rKN{16}
\def\rKNS{17}
\def\rJKMO{18}
\def\rP{19}
\def\rDr{20}
\def\rJi{21}
\def\rRW{22}
\def\rRe{23}
\def\rORW{24}
\def\rYY{25}
\def\rTak{26}
\def\rGa{27}
\def\rTS{28}
\def\rBab{29}
\def\rTakh{30}
\def\rBDV{31}
\def\rAD{32}
\def\rJSu{33}
\def\rABF{34}
\def\rJMO{35}
\def\rKS{36}
\def\rKuO{37}
\def\rWSN{38}
\def\rDJKMO{39}
\def\rKR{40}
\def\rGKO{41}
\def\rKNlev{42}
\def\rKNC{43}
\def\rNRT{44}
\def\rDKMM{45}
\def\rFS{46}
\def\rDuSa{47}
\def\rGe{48}
\def\rKRjp{49}
\def\rKPet{50}
%
%
%
%
\def\reference{
\vfill\eject
\vskip0.6cm\pn
{\bf References}
\item{[\rPaI]}{A.Kuniba, T.Nakanishi and J.Suzuki,
  {\sl Functional Relations in
  Solvable Lattice Models I: Functional relations and
Representation theory,} hep-th.9309137, HUTP-93 A022.}
\item{[\rBax]}{R.\ J.\ Baxter, {\sl Exactly Solved Models in Statistical
Mechanics}
      ,(Academic Press, London 1982).}
\item{[\rBP]}{R.\ J.\ Baxter and P.\ A.\ Pearce,
       J.\ Phys. A. Math. Gen. {\bf 15} (1982) 897; {\bf 15} (1983) 2239.}
\item{[\rKBP]}{A.\ Kl{\"u}mper, M.\ T.\ Batchelor and P.\ A.\ Pearce,
   J.\ Phys. A. Math. Gen. {\bf 24} (1991) 3111.}
\item{[\rKSZ]}{A.\ Kl{\"u}mper, A.\ Schadshneider and J.\ Zittarz,
       Z.\ Phys.\ {\bf B76} (1989)247.}
\item{[\rKPO]}{A.\ Kl{\"u}mper and P.\ A.\ Pearce,
    Phys.Rev.Lett {\bf 66} (1991) 974;
    J.\ Stat.\ Phys {\bf 64} (1991)13.}
\item{[\rKPT]}{A.\ Kl{\"u}mper and P.\ A.\ Pearce,
    Physica {\bf A183} (1992) 304.}
\item{[\rJKM]}{J.\ D.\ Johnson, S.\  Krinsky and B.\ M.\  McCoy,
      Phys.\  Rev.\  {\bf A8} (1973) 2526.}
\item{[\rBCN]}{H.\ W.\ Bl{\"o}te, J.\ L.\ Cardy and M.\ P.\ Nightingale,
   Phys.\ Rev.\ Lett.\ {\bf 56} (1986) 742.}
\item{[\rAf]}{I.\ Affleck,
   Phys.\ Rev.\ Lett.\ {\bf 56} (1986)746.}
\item{[\rKi]}{A.\ N.\ Kirillov, J.\ Sov.\ Math.{\bf 47} (1989) 2450.}
\item{[\rKu]}{A.\ Kuniba, Nucl.\ Phys.\ {\bf B389} (1993) 209.}
\item{[\rBRO]}{V.\ V.\ Bazhanov and N.\ Yu.\ Reshetikhin,
        Int.\ J.\ Mod.\ Phys.\ {\bf A4} (1989) 115.}
\item{[\rBRT]}{V.\ V.\ Bazhanov and N.\ Yu.\ Reshetikhin,
        J.\  Phys. A. Math. Gen. {\bf 23} (1990) 1477.}
\item{[\rAM]}{F.\ C.\ Alcaraz and M.\ J.\ Martins,
    J.\ Phys. A. Math. Gen. {\bf 21} (1988) 4397.}
\item{[\rKN]}{A.\ Kuniba and T.\ Nakanishi,
    Mod.\ Phys.\ Lett {\bf A7} (1992) 3487.}
\item{[\rKNS]}{A.\ Kuniba, T.\ Nakanishi and J.\ Suzuki,
    Mod.\ Phys.\ Lett {\bf A8} (1993) 1649}
\item{[\rJKMO]}{M.\ Jimbo, A.\ Kuniba, T.\ Miwa, M.\ Okado,
Commun. \ Math.\ Phys.\ {\bf 119} (1988) 543.}
\item{[\rP]}{P.\ A.\ Pearce, Phys. Rev. Lett {\bf 58} (1987) 1502.}
\item{[\rDr]}{V.\ G.\ Drinfel'd, Sov.\ Math.\ Dokl.\  {\bf 32} (1985) 254;
in {\it ICM Proceedings, Berkeley},
(American Mathematical Society, 1987)}
\item{[\rJi]}{M.\ Jimbo, Lett.\ Math.\ Phys.{\bf 10} (1985) 537.}
\item{[\rRW]}{N.\ Yu.\ Reshetikhin and P.\ Wiegmann,
     Phys.\  Lett.\ {\bf B189} (1987) 125.}
\item{[\rRe]}{N.\ Yu.\ Reshetikhin, Lett.\ Math.\ Phys.\ 14 (1987) 235.}
\item{[\rORW]}{E.\ Ogievetsky, N.\ Yu.\ Reshetikhin and P.\ Wiegmann,
    Nucl.\ Phys.\ {\bf B280} [FS18] (1987) 45.}
\item{[\rYY]}{C.\ N.\ Yang and C.\ P.\ Yang,
      J.\ Math.\ Phys.\ {\bf 10} (1969) 1115.}
\item{[\rTak]}{M. Takahashi, Phys. Lett. {\bf A36} (1971) 325.}
\item{[\rGa]}{M. Gaudin, Phys. Rev. Lett. {\bf 26} (1971) 1301.}
\item{[\rTS]}{M. Takahashi and M. Suzuki, Prog. Theor. Phys. {\bf 48} (1972)
2187.}
\item{[\rBab]}{M.\ Babujian,  Nucl.\ Phys.\ {\bf B125} (1983) 317.}
\item{[\rTakh]}{L.\ A.\ Takhatajian, Phys.\ Lett.\ {\bf A87} (1982) 479.}
\item{[\rBDV]}{O.\ Babelon, H.\ J.\ de Vega and C.\ M.\  Viallet,
      Nucl.\ Phys.\ {\bf B220} (1983) 13.}
\item{[\rAD]}{L.\ V.\ Avdeev and B.\ D.\ Dorfel,
       Nucl.\ Phys.\ {\bf B257} (1985) 253.}
\item{[\rJSu]}{J.\ Suzuki, J.\ Phys.\ Soc.\ Jpn.\ {\bf 58} (1989) 3111.}
\item{[\rABF]}{G.\ E.\ Andrews, R.\ J.\ Baxter and P.\ J.\ Forrester,
      J.\ Stat.\ Phys.\ {\bf 35} (1984) 193.}
\item{[\rJMO]}{M.\ Jimbo, T.\ Miwa, M.\ Okado,
\ Commun.\ Math.\ Phys.\ {\bf 116} (1988) 507.}
\item{[\rKS]}{A.\ Kuniba and J.\ Suzuki,
 Phys.\ Lett.\ {\bf A166} (1991) 216.}
\item{[\rKuO]}{A.\ Kuniba, Nucl.\ Phys.\  {\bf B355} (1991) 801.}
\item{[\rWSN]}{S.\ O.\ Warnaar, K.\ A.\ Seaton and B.\ Nienhuis,
   Phys.\ Rev.\ Lett.\ {\bf 69} (1992) 710.}
\item{[\rDJKMO]}{E.\ Date, M.\ Jimbo, A.\ Kuniba, T.\ Miwa, M.\ Okado,
 Nucl.\ Phys.\ {\bf B290} [FS20]
  (1987) 231; Adv.\ Stud.\ in Pure Math.\ {\bf 16} (1988) 17.}
\item{[\rKR]}{A. N.\ Kirillov and N. Yu.\ Reshetikhin,
Zap.\ Nauch.\ Semin.\ LOMI {\bf 160} (1987) 211
[J.\ Sov.\ Math.\ {\bf 52} (1990) 3156]}
\item{[\rGKO]}{P.\ Goddard, A.\ Kent and D.\ Olive
  Phys.\ Lett. {\bf B152} (1985) 88.}
\item{[\rKNlev]}{A.\ Kuniba and T.\ Nakanishi, in {\it Modern Qunatum
Field Theory}, ed. S. Das, A. Dhar, S. Mukhi, A. Raina and A. Sen,
(World Scientific, Singapore, 1991).}
\item{[\rKNC]}{A. Kuniba and T. Nakanishi, in
{\it proceedings of the
XXI th Differential Geometric Methods in Theoretical Physics},
ed. C.\ N.\ Yang, M.\ L.\ Ge and X.\ W.\ Zhou
(World Scientific, Singapore, 1993).}
\item{[\rNRT]}{W.\ Nahm, A.\ Recknagel and M.\ Terhoeven,
{\sl Dilogarithm identities in conformal field theory},
preprint, BONN-HE-92-35.}
\item{[\rDKMM]}{S.\ Dasmahapatra, R.\ Kedem, T.\ R.\ Klassen,
B.\ M.\  McCoy and E.\ Melzer, {\sl Quasi-Particles, Conformal Field Theory and
q-Series}, prerpint, hep-th/9303013.}
\item{[\rFS]}{E.\ Frenkel and A.\ Szenes,
Int.\ Math.\ Res.\ Notices {\bf 2 }(1993) 53;
{\sl Crystal Bases, Dilogarithm Identities
and Torsion in Algebraic K-Group}, preprint.}
\item{[\rDuSa]}{J.\ L.\ Dupont and C.\ H.\ Sah,
"Dilogarithm Identities in Conformal Field Theory
and Group Homology", hep-th/9303111.}
\item{[\rGe]}{D.\ Gepner, Nucl.\ Phys.\ {\bf B290} [FS20] (1987) 10.}
\item{[\rKRjp]}{A.\ N.\ Kirillov and  N.\ Yu.\ Reshetikhin,
   J.\ Phys. A. Math. Gen. {\bf 20} (1987) 1587.}
\item{[\rKPet]}{V.\ G.\ Kac and D.\ H.\ Peterson,
   Adv.\ Math.\ {\bf 53}(1984) 125.}
}
\Title{HUTP-93/A023}
{\vbox{\centerline{Functional Relations
       in Solvable Lattice Models II: }
        \vskip3pt
       \centerline{Applications}}}
\centerline{Atsuo Kuniba\footnote{$^1$}
{e-mail: kuniba@math.sci.kyushu-u.ac.jp}
, Tomoki Nakanishi\footnote{$^2$}
{e-mail: nakanisi@string.harvard.edu}
\footnote{\null}{Permanent address:
Department of Mathematics, Nagoya University,
Nagoya 464 Japan}and  Junji
Suzuki\footnote{$^3$}
{e-mail: jsuzuki@tansei.cc.u-tokyo.ac.jp}}
\vskip0.5cm
\bigskip\centerline{
$^1$Department of Mathematics, Kyushu University,
Fukuoka 812 JAPAN}
\bigskip\centerline{
$^2$Lyman Laboratory of Physics, Harvard University,
Cambridge, MA 02138 USA}
\bigskip\centerline{
$^3$Institute of Physics, University of Tokyo, Komaba,
Meguro-ku, Tokyo 153 JAPAN}

\vskip .3in
%
%
Abstract. Reported are two applications of the functional relations
($T$-system)  among a commuting family of row-to-row transfer matrices
proposed in the previous paper Part I.
For a general simple Lie algebra $X_r$, we determine
the correlation lengths of the associated massive vertex models in the
anti-ferroelectric regime
and central charges of the RSOS models in two critical regimes.
The results reproduce known values or even generalize them,
demonstrating the efficiency of the $T$-system.

\Date{10/93}
%
%
%
%
\sect{\bf{1. Introduction}}
\subskip

This is a continuation of our previous paper, hereafter called Part I [\rPaI].
There, we have proposed the functional
relation (FR), the $T$-system, among a commuting family of
row-to-row transfer matrices for a class of solvable lattice models.
They are the vertex and RSOS type models
associated with any simple Lie algebra $X_r$
in the sense of section 3.3 of Part I.
Let $\{T^{(a)}_m(u) \}$ be the family of their row-to-row transfer matrices
acting on a common quantum space.
Here $u$ denotes the spectral parameter and
the $(a,m)$ labels the fusion type, i.e.,
signifies that the
auxiliary space is the
$U_q(X^{(1)}_r)$-module $W^{(a)}_m(u)$.
See section 3.2 and Fig.2 in Part I.
Then the $T$-system is the following three term FRs for
$a=1,\cdots, r$, $m=1,2,\ldots$;
$$
T^{(a)}_m(u-{1 \over {2t_a}})T^{(a)}_m(u+{1\over {2t_a}})=
T^{(a)}_{m-1}(u)T^{(a)}_{m+1}(u) +
g^{(a)}_m(u) \prod_{b=1}^{r} {\cal T}(a,b,m,u)^{I_{ab}},
\eqno(1.1{\rm a})
$$
where the explicit forms of
${\cal T}(a,b,m,u)$ are given by
$$\eqalignno{
{\cal T}(a,b,m,u) =&T^{(b)}_{t_b m /t_a}(u)
                   \qquad {\hbox{for }}  {t_b \over t_a}=1,2,3, &(1.1{\rm
b})\cr
                  =&T^{(b)}_{m\over 2}(u-{1\over 4})
                    T^{(b)}_{m\over 2}(u+{1 \over 4})\cr
                   &\times T^{(b)}_{m-1\over 2}(u)T^{(b)}_{m+1\over 2}(u)
                  \qquad {\hbox{for }} {t_b \over t_a}={1 \over 2},
                                       &(1.1{\rm c}) \cr
       =&T^{(b)}_{m\over 3}(u-{1 \over3})T^{(b)}_{m\over 3}(u)
        T^{(b)}_{m\over 3}(u+{1 \over3})    \cr
        &\times T^{(b)}_{m-1\over 3}(u-{1\over 6})
        T^{(b)}_{m-1\over 3}(u+{1\over 6}) T^{(b)}_{m+2\over 3}(u)\cr
        &\times T^{(b)}_{m+1\over 3}(u-{1\over 6})
        T^{(b)}_{m+1\over 3}(u+{1\over 6}) T^{(b)}_{m-2\over 3}(u)
          \,{\hbox{ for }} {t_b \over t_a}={1 \over 3}, &(1.1{\rm d})
}$$
under the conventions
$\forall T^{(a)}_0(u) = 1$ and
$T^{(a)}_m(u) = 1$ if $m \not\in {\bf Z}$.
In (1.1a), $g^{(a)}_m(u)$ is a scalar function
depending on the quantum space.
The integers $t_a \in \{1,2,3\}$ and $I_{ab}\in \{0,1\}$
are defined from the root system of $X_r$ (cf. Part I, eq.(3.1)):
$$\eqalign{
I_{a b}&=2 \delta_{ab}-B_{a b}, \cr
B_{a b}&=C_{a b} {{t_b }\over{ t_{ab}}},
\qquad C_{a b} = {2(\alpha_a \vert \alpha_b) \over
(\alpha_a \vert \alpha_a)},\cr
t_a&={2 \over {(\alpha_a|\alpha_a)}}, \qquad
t_{ab}=\hbox{max}(t_a,t_b).}
\eqno(1.2)$$
We call (1.1) the {\it unrestricted} $T$-system
when we let $m$ extend over all positive integers.
On the other hand,
if $T^{(a)}_{t_a\ell+1}(u) \equiv 0 \,(1 \le a \le r)$
is imposed for some integer $\ell \ge 1$, (1.1) closes among
those $T^{(a)}_m(u)$'s with $1 \le a \le r, 1 \le m \le t_a\ell$.
We call it the (level $\ell$) {\it restricted} $T$-system.
These are to hold for the vertex and the level $\ell$ RSOS models
associated with $U_q(X^{(1)}_r)$, respectively.
\par
Applications of the above $T$-system
to the calculation of physical quantities
are the main purpose in this report Part II.
More specifically,
we will determine
the correlation lengths of massive vertex
models in anti-ferroelectric regime (section 2) and
central charges of critical RSOS models (section 3).
The idea of using the FRs
for computing thermodynamic
quantities is not itself new but
often known to bypass complicated
Bethe ansatz analyses (cf.[\rBax,\rBP,\rKBP]).
Our $T$-system approach indeed possesses such an advantage and generalizes
the earlier studies [\rKSZ-\rKPT] for the simplest
$X_r = A_1$ case.
\par
As for the correlation lengths of the
massive vertex models,
the derivation reduces to finding a periodicity
of $T^{(a)}_m(u)$ that obeys a truncated version (2.6) of (1.1a).
We call it the  ``bulk" $T$-system and
prove that the real period is just the dual Coxeter number,
thereby giving a
unified expression (2.3) for the correlation length.
This extends the earlier analysis for the $sl(2)$ case in [\rKSZ].
As an independent check,
we shall also invoke the standard Bethe ansatz method [\rJKM]
and show the agreement of the results.
\par
To evaluate the central charges, we
generalize the treatment of the $sl(2)$ case [\rKPO,\rKPT]
to arbitrary $X_r$.
We convert the $T$-system into integral equations for
finite size corrections and extract
the central charges
by the standard argument [\rBCN,\rAf].
It leads to the expressions (3.29) and (3.49)
in terms of the Rogers dilogarithm, which turn out to be calculable by
means of the conjecture in [\rKi,\rKu].
The results reproduce those in [\rKu-\rBRT]
obtained from the thermodynamic Bethe ansatz (TBA) analyses.\par
These agreements show the efficiency of
our $T$-system approach.
It works well even for
the models that defy the root density method [\rAM]
and leads to scaling dimensions
as shown in [\rKPO,\rKPT] for $sl(2)$.
Though we only determine the central charges in section 3,
our formulation also covers the scaling dimensions,
for which the relevant dilogarithm conjecture is
now available in [\rKN].
We hope to further extend the
analyses presented in this paper to actually derive the
scaling dimensions or even to produce the character itself
for the corresponding conformal field theories (CFTs)
(cf. [\rKNS]).\par
Appendix A supplements the proof of the periodicity
of the bulk $T$-system.
Appendix B provides a list of the matrix functions
needed in the Bethe ansatz computation in section 2.3.
Appendix C is a calculation of a certain ratio of the bulk eigenvalues of the
transfer matrices [\rBRT] for the $A^{(1)}_r$ RSOS models [\rJKMO].
It is an important input to the integral equations in section 3.
Appendix D is devoted to an exposition of the generalized dilogarithm
conjecture in [\rKN].
Throughout the paper, the equations in Part I [\rPaI], e.g., (B.3), will
be referred to as (IB.3), etc.
\secskip
%
%
%
%
\centerline{\bf{ 2. Correlation Lengths of Massive Vertex Models}}\pn
\subskip
As the first application of our $T$-system (1.1),
we shall compute the correlation lengths of the massive vertex models
in anti-ferroelectric regime.
The same quantities will be calculated through a
Bethe ansatz method as an independent check.
\subskip\noindent
{\bf 2.1. Massive vertex models in anti-ferroelectric regime}\par
Let us briefly sketch what we mean by massive vertex models
in anti-ferroelectric (AF) regime.
As the simplest example, consider the 6-vertex model
with the Boltzmann weights given by
$$\eqalign{
w(\pm 1, \pm 1, \pm 1, \pm 1) &= {\sinh \lambda(1-u)\over \sinh \lambda},\,
w(\pm 1, \mp 1, \pm 1, \mp 1) = {\sinh \lambda u\over \sinh \lambda},\cr
w(\pm 1, \pm 1, \mp 1, \mp 1) &= 1,\cr}
\eqno(2.1)
$$
where the variables $\epsilon_1, \ldots, \epsilon_4$ in
$w(\epsilon_1, \ldots, \epsilon_4)$ is ordered clockwise
from the left edge of the vertex.
We impose the periodic boundary condition and
assume that the system size is even.
In (2.1) one observes that\vskip0.1cm\pn
\item{(i)} the Boltzmann weights are trigonometric
functions of $u$ and $\lambda$ and
are analytic with respect to $u$.
The partition function with periodic boundaries is invariant under
$u \rightarrow u+{\pi i\over\lambda}$.\vskip0.1cm\pn
The invariance can be seen by noting that
the weights $w(1,1,-1,-1)$ and
$w(-1,-1,1,1)$ always occur in pairs.
We shall consider the regime $0 < u < {1 \over 2},\, \lambda > 0$.
Then (2.1) corresponds to
$\Delta <-1$ in the conventional parameter [\rBax]
and it is well known that
\vskip0.1cm\pn
\item{(ii)} the model is critical as $\lambda \rightarrow 0$
and is in an
AF ground state as
$\lambda \rightarrow \infty$.\vskip0.1cm\pn
This is natural in view of that
all the weights (2.1) compete when
$\lambda \rightarrow 0$ while only the last one dominates
exponentially as $\lambda \rightarrow \infty$.
The ground state configurations are invariant in the
SW-NE direction and alternating in the NW-SE direction.
Thus the range $0 < \lambda < \infty$ corresponds to
the AF phase.
As we saw in section 2 of Part I, one can form a length $N$
row-to-row transfer matrix $T_1(u)$ acting in the
vertical direction and obeying the $T$-system
$$
T_1(u-{1 \over 2})T_1(u+{1 \over 2}) = T_2(u) +  g_2^N(u) {\hbox{Id}}
\eqno(2.2)
$$
with $g_2(u)=\sinh\lambda(3/2-u)\sinh\lambda(1/2+u)/\sinh^2\lambda$.
Here $T_2(u)$ is the transfer matrix for the degree 2 fusion model
in the auxiliary space.
(The spectral parameter in section 2 of Part I is denoted by $-u$ here.)
When evaluating (2.2)
on the common eigenvectors, it is generally expected
in the regime $0 < u < {1 \over 2},\, \lambda > 0$ [\rBP,\rP] that
\vskip0.1cm\pn
\item{(iii)}
${\vert\hbox{eigenvalue of the 1st term}\vert
\over
\vert\hbox{eigenvalue of the 2nd term}\vert} < e^{-\delta N}$
with some $\delta >0$ on the rhs of the $T$-system,
\vskip0.1cm\pn
which simplifies (2.2) in the thermodynamic limit
$N \rightarrow \infty$.
Excitations are described by various eigenstates of the transfer matrices.
Then,
\vskip0.1cm\pn
\item{(iv)}
there are finitely many excited states that degenerate
to the lowest one as
$N \rightarrow \infty$.\vskip0.1cm\pn
We shall call all these finitely many states as the first
excited states.
In the present 6-vertex case, there is only one first excited
state relevant to the interfacial tension.
The lowest excitations above the first ones are called
the second excitations.
Then it is known that
\vskip0.1cm\pn
\item{(v)}
the second excited states form an energy band for
$N \rightarrow \infty$ and they are higher than the first excitation
energy by a certain gap,\vskip0.1cm\pn
which implies that the model is massive and the correlation length
is just the inverse of the energy gap.
Note that this defines the vertical correlation length
which should be the same for all the fusion models
whose transfer matrices act on the common quantum space.
\par
Now recall that
the 6-vertex model (2.1) is associated to
the fundamental representation of the quantum group
$U_q(A^{(1)}_1)$ with $q=e^{-\lambda}$ [\rDr,\rJi].
Analogously we consider the fusion vertex models
associated to $U_q(X^{(1)}_r)$ with the fixed
homogeneous quantum space $W^{(p)}_s{}^{\otimes N}$
for some integers $1 \le p \le r$ and $s \ge 1$.
See section 3 of Part I.
As discussed there, they are expected to obey the
unrestricted $T$-system (1.1).
Moreover we suppose (i) and that there exists a
certain parameter regime in $u$ and $\lambda$
such that all the features (ii)-(v) are valid.
We call them the massive $U_q(X^{(1)}_r)$ vertex models in AF regime.
\par
In the rest of this section we exclude the
case $X_r = A_r$ with $r \ge 2$ and take the AF regime
as $0 < u < {1 \over 2}, \lambda > 0$.
We will show that the
correlation length $\xi$ of these vertex models
is given by a unified formula
$$
\xi = - {1 \over \log k}. \eqno(2.3{\rm a})
$$
Here $0 < k < 1$ is determined by
$$
{K^\prime(k) \over K(k)} = {g\lambda \over \pi},\eqno(2.3{\rm b})
$$
where $g$ is the dual Coxeter number of $X_r$ and
$K(k) \, (K^\prime(k))$
denotes the complete elliptic integral of the first (second) kind with
modulus $k$ (cf. section 15 in [\rBax]).
Namely,
$$
k = 4y^{1\over 2}\prod_{n=1}^\infty\Bigl(
{1+y^{2n}\over 1+y^{2n-1}}\Bigr)^4\quad  y = e^{-g\lambda}.
\eqno(2.3{\rm c})
$$
Note that (2.3) is consistent with the property (ii):
 $\xi \rightarrow 0(\infty)$ as $\lambda \rightarrow \infty (0)$.
Actually it reproduces the 6-vertex model result
eq.(8.10.12) in [\rBax] for $X_r = A_1\, (g=2)$
up to a trivial overall factor
which depends on the normalization of $\xi$.
Such factors will not be concerned hereafter.
The result (2.3) will be shown by two independent methods.
One uses our $T$-systems and the other is a rather conventional
Bethe ansatz analysis where we will use a string
hypothesis (vi) (see section 2.3) as well.
Both methods lead to (2.3),
supporting our hypotheses on the $T$-system.
(In the Bethe ansatz analysis,
we shall actually consider a less
general setting
as will be noted in the beginning of section 2.3.)\par
In the working below
the AF regime will always be taken as $0 < u < {1 \over 2}$.
This is actually needed in the derivation using the Bethe ansatz
in section 2.3.
See a remark before (2.42).
However, the analyses indicate that (2.3) is valid in a wider range
$0 < u < {g \over 2}$ ($g$: dual Coxeter number).
This has been observed for the
6-vertex model in p155 of [\rBax] and will also be argued
in the end of section 2.3.
\subskip
%
%
\noindent
{\bf 2.2. Correlation length from the $T$-system}
\par\noindent
{\it 2.2.1 Double Periodicity} \par
Consider the $T$-system (1.1) among the
length $N$ transfer matrices $T^{(a)}_m(u)$
for the massive vertex models.
We denote by $L^{(a)}_m(u)$ the
ratio of the eigenvalues of $T^{(a)}_m(u)$ for
a second excitation to the ground
state.
Then $L^{(a)}_m(u)$ turns out to be
a doubly periodic function with respect to $u$ in the thermodynamic limit,
which is crucial in deriving the correlation lengths [\rKSZ].
{}From the property (i),  $L^{(a)}_m(u)$
is a meromorphic function of $u$ and has the periodicity in the
imaginary direction of $u$,
$$L^{(a)}_m(u) = L^{(a)}_m(u + {\pi i \over \lambda}).
\eqno(2.4)
$$
We set $N\rightarrow \infty$ from now on and
seek the periodicity in the real direction.
We shall exclusively consider the case $m\in t_a {\bf Z}_{\ge 0}$
in the sequel.
In this limit, $L^{(a)}_m(u)$ actually measures the energy gap
due to the property (iv).
Let us quote the unrestricted $T$-system (1.1a)
symbolically as a three term relation
$${\cal T}_0 = {\cal T}_1 + {\cal T}_{-1},
\eqno(2.5)
$$
where ${\cal T}_0 = T^{(a)}_m(u + {1\over {2 t_a}})
T^{(a)}_m(u-{1\over {2 t_a}})$,
${\cal T}_1 = T^{(a)}_{m+1}(u) T^{(a)}_{m-1}(u)$
and the third term ${\cal T}_{-1}$ is also a product of
$T^{(a^\prime)}_{m^\prime}$'s and the scalar function $g^{(a)}_m(u)$.
We assume that
$\vert {\cal T}_1 \vert \ll \vert {\cal T}_{-1} \vert$
as $N \rightarrow \infty$ in AF regime.
This is a natural extension of the property (iii) for the $sl(2)$ case [\rP]
where it is consistent with the fact that
the ground state is singlet.
In this view, we introduce the truncated version of (2.5)
$${\cal T}_0 = {\cal T}_{-1}\vert_{\forall g^{(a)}_m(u)\equiv 1}, \eqno(2.6)
$$
which we call the bulk $T$-system.
By the definition, $L^{(a)}_m(u)$ satisfies (2.6) because
the $g^{(a)}_m(u)$
factor cancells when taking the ratio of the second excitation
to the ground state.
Then, the point is that the bulk $T$-system itself imposes the periodicity
as follows.
\proclaim Proposition.
Suppose $T^{(a)}_m(u)$ satisfies the bulk $T$-system (2.6).
Then the following periodicity is valid for $m \in t_a{\bf Z}_{\ge 0}$.
$$\eqalignno{
&T^{(a)}_{ m}(u) = T^{(a)}_{ m}(u + g)\quad \hbox{ for any } \, X_r,
&(2.7)\cr
&{\bar T}^{(a)}_{ m}(u){\bar T}^{(a)}_{ m}(u + {g \over 2}) = 1
\quad \hbox{ if }\,\, X_r \ne A_{r \ge 2},&(2.8{\rm a})\cr
&{\bar T}^{(a)}_{ m}(u) = \cases{
T^{(a)}_{ m}(u)T^{(a)}_{ m}(u+1)T^{(a)}_{ m}(u+2)
& if $X_r = E_6$ or $E_8$,\cr
T^{(a)}_{ m}(u) & if $X_r \ne A_{r \ge 2}, E_6$ or $E_8$,\cr}
&(2.8{\rm b})\cr}
$$
where $g$ denotes the dual Coxeter number.\pn

Here we have also included the case $X_r = A_{r \ge 2}$ in the
statement
although it is irrelevant to the correlation length calculation.
Note that (2.7) is just a corollary of (2.8) if
$X_r \ne A_{r \ge 2}, E_6$ or $E_8$.
The proposition claims another period than (2.4) as
$$
L^{(a)}_m(u) =L^{(a)}_m(u+g)\quad \hbox{ for }
\quad m\in t_a {\bf Z}_{\ge 0}.
\eqno(2.9)
$$
This is consistent with $Y$-system's period (IB.7)
in view of the connection (I3.19) and the
fact that the bulk $T$-system formally corresponds
to the level 0 restricted $T$-system.
In the rest of this subsection,
we illustrate the proof of the proposition for $X_r=A_r, B_r$
and leave the other cases to appendix A.\pn \vskip0.3cm
\pn
{\it Proof.}\hskip0.2cm
\noindent $X_r=A_r$:
The bulk $T$-system is
$$
T^{(a)}_m(u-\hlf) T^{(a)}_m(u+\hlf) =T^{(a+1)}_m(u) T^{(a-1)}_m(u)
\quad 1 \le a \le r,
\eqno(2.10)
$$
where we have put $T^{(0)}_m(u)=T^{(r+1)}_m(u)=1$.
By induction, it is easy to show
$$
T^{(a)}_m(u) = \prod_{j=1}^{a}T^{(1)}_m(u+j-{{a+1}\over 2}),
\eqno(2.11)
$$
for $1 \le a \le r+1$.
Setting $a=r+1$ and $u=u'+r/2$ in the above, we obtain
$$
T^{(1)}_m(u') \cdots T^{(1)}_m(u'+r)=1,
\eqno(2.12)
$$
from which $T^{(1)}_m(u)=T^{(1)}_m(u+r+1)$ hence (2.7) follows.
Note that eq.(2.8a) holds only for $r=1$ as clearly seen from the
above derivation.
\vskip0.2cm\pn
$X_r=B_r$:
The relevant bulk $T$-system is
$$\eqalignno{
T^{(a)}_m(u-\hlf) T^{(a)}_m(u+\hlf) &= T^{(a+1)}_m(u) T^{(a-1)}_m(u)
\quad \hbox{ for }\,\, 1 \le a \le r-2,\cr
&&(2.13{\rm a})\cr
T^{(r-1)}_m(u-\hlf) T^{(r-1)}_m(u+\hlf) &=T^{(r-2)}_m(u) T^{(r)}_{2m}(u),
       &(2.13{\rm b})    \cr
T^{(r)}_{2m}(u-\qtr) T^{(r)}_{2m}(u+\qtr)
       &= T^{(r-1)}_m(u-\qtr) T^{(r-1)}_{m}(u+\qtr),
&(2.13{\rm c})\cr
}$$
where $T^{(0)}_m(u) = 1$.
{}From (2.13a) we have
$$
T^{(a)}_m(u)=\prod_{j=1}^{a}T^{(1)}_m(u+j-{{a+1}\over 2}),
\eqno(2.14)
$$
for $1 \le a \le r-1$.
Using this in (2.13b), we find
$$
T^{(r)}_{2m}(u) = \prod_{j=1}^{r}T^{(1)}_m(u+j-{{r+1}\over 2}),
\eqno(2.15)
$$
Substitute (2.14) and (2.15) into (2.13c) and cancel the common factors.
The result reads
$$
T^{(1)}_m(u-{{2r-1}\over 4}) T^{(1)}_m(u+{{2r-1}\over 4})=1.
\eqno(2.16)
$$
This establishes (2.8) for all $1 \le a \le r$
because of (2.14,15) and $g=2r-1$.\pn
\subskip\noindent
{\it 2.2.2 Correlation Lengths from the double periodicity} \par
So far we have seen that $L^{(a)}_m(u)$ is a meromorphic
and doubly periodic function of $u$ as specified in (2.4) and (2.9).
Next we define $0 < k < 1$ by (2.3b,c) and put
$$\eqalignno{
h_1(u,u_0)&=\sqrt{k} \hbox{ sn}\Bigl(
{2i\lambda K(k)\over \pi}(u-u_0)\Bigr),&(2.17{\rm a})\cr
h_2(u,u_0)&=\sqrt{k} \hbox{ sn}\Bigl(
{2i\lambda K(k)\over \pi}(u-u_0+{g \over 2})\Bigr),&(2.17{\rm b})\cr}
$$
where Jacobi's elliptic function $\hbox{sn}$ is of modulus $k$.
These are meromorphic, $g$-periodic,
${\pi i \over \lambda}$-anti-periodic functions of $u$ and
satisfy
$$
h_j(u,u_0)h_j(u+{g\over 2},u_0) = 1\quad \hbox{for } j = 1,2.
\eqno(2.18)
$$
In the rectangle
$[0,{g\over 2})\times [0,{\pi i \over \lambda})$ on the
complex $u$-plane, $h_1(u,u_0)$ $ (h_2(u,u_0))$
has one simple zero (pole) and no poles (zeros).
Put
$$
{\bar L}^{(a)}_m(u) = \cases{
L^{(a)}_m(u)L^{(a)}_m(u+1)L^{(a)}_m(u+2) & if $X_r = E_6$ or $E_8$,\cr
L^{(a)}_m(u) & otherwise,\cr}
\eqno(2.19)
$$
and let $\{u_z\},\, \{u_p \}$ be the set of zeros and poles of
${\bar L}^{(a)}_m(u)$ in the rectangle
$u \in [0,{g\over 2})\times [0,{\pi i \over \lambda})$, respectively.
Then the ratio
$$h(u) =
{{\bar L}^{(a)}_m(u) \over
\prod_{u_z}h_1(u,u_z)
\prod_{u_p}h_2(u,u_p)}\eqno(2.20)
$$
is analytic and non-zero for $0 \le Re(u) < {g\over 2}$.
Furthermore from (2.8) we have
$$h(u)h(u+{g\over 2}) = 1\eqno(2.21)$$
except for $X_r=A_{r \ge 2}$.
Therefore $h(u)$ is doubly periodic with periods
$g$ and ${\pi i \over \lambda}$
(if $\#\{u_z\} + \#\{u_p\}$ is even) and it is
analytic and non-zero on the whole complex $u$-plane.
{}From the Liouville theorem and (2.21) it follows that
$h(u) = \pm 1$.
Thus we obtain
$$
{\bar L}^{(a)}_m(u) = \pm
\prod_{u_z}\sqrt{k} \hbox{ sn}\Bigl(
{2i\lambda K(k)\over \pi}(u-u_z)\Bigr)
\prod_{u_p}\sqrt{k} \hbox{ sn}\Bigl(
{2i\lambda K(k)\over \pi}(u-u_p+{g\over 2})\Bigr),
\eqno(2.22)
$$
where there must be even number of $\hbox{sn}$'s in total on the rhs for
${\bar L}^{(a)}_m(u)$ to be ${\pi i \over \lambda}$-periodic.
\par
Consider now the correlation function
$G(R) = \langle \delta_{\sigma_0 \rho_1}\delta_{\sigma_R \rho_2}
\rangle$ of the two vertical edge variables
$\sigma_0,\,\sigma_R$ on the same column and separated by
$R$ vertices.
Here $\rho_1, \, \rho_2$ denote some given edge states.
The vertical correlation length $\xi$
is then defined by
$$G(R) - G(\infty) \simeq \hbox{const} \cdot e^{-{R\over \xi}}\quad
\hbox{ as } R \rightarrow \infty.\eqno(2.23)
$$
To extract the $\xi$ from (2.22) and (2.23)
we follow [\rKSZ,\rJKM] and need a few more
assumptions.
Firstly, we assume that ${\bar L}^{(a)}_m(u)$ is free of poles
in $0 \le Re(u) \le {g \over 2}$.
This may be natural since its denominator is the
ground state eigenvalue of $T^{(a)}_m(u)$.
Secondly, the leading part of $G(R) - G(\infty)$ will
come from those ${\bar L}^{(a)}_m(u)$ that contain
only two zeros.
Thus the relevant case of (2.22) has the from
$$
{\bar L}^{(a)}_m(u) = {\bar L}^{(a)}_m(u; u_1,u_2) = \pm k
\hbox{ sn}\Bigl(
{2i\lambda K(k)\over \pi}(u-u_1)\Bigr)
\hbox{ sn}\Bigl(
{2i\lambda K(k)\over \pi}(u-u_2)\Bigr).
\eqno(2.24)
$$
Various choices for $u_1$ and $u_2$ here will correspond
to the energy band mentioned in the property (v).
The correlation function is now evaluated by assembling the
contribution from the band as
$$
G(R) - G(\infty) \simeq \int du_1 \int du_2 \, g(u_1,u_2)
\Bigl({\bar L}^{(a)}_m(u; u_1,u_2)\Bigr)^R,
\eqno(2.25)
$$
where $g(u_1,u_2)$ is some weight function and the integration is
over a finite range.
Finally, we assume that the integral
$\int du_1 \int du_2 \, g(u_1,u_2)
\hbox{ sn}^R\bigl(
{2i\lambda K(k)\over \pi}(u-u_1)\bigr)
\hbox{ sn}^R\bigl(
{2i\lambda K(k)\over \pi}(u-u_2)\bigr)
$
does not develop a dominant $R$-dependence
in the limit $R \rightarrow \infty$ compared
with $k^R$.
The discrepancy between
${\bar L}^{(a)}_m$ and $L^{(a)}_m$
for $X_r = E_6$ and $E_8$ is expected to be
irrelevant to extracting the dominant
$R$-dependence.
Thus substitution of (2.24) into (2.25) leads to
$$G(R) - G(\infty) \simeq \hbox{const}\cdot k^R.$$
Comparing this with (2.23) we arrive at (2.3).\par
%
%
\subskip\noindent
{\bf 2.3. Correlation length from Bethe ansatz}
\pn
{\it 2.3.1 Explicit forms of the free energy and the Bethe ansatz
   equation} \par
The Bethe ansatz equation for our
$U_q(X^{(1)}_r)$ massive vertex model
has the form $(1 \le a \le r)$ [\rRW-\rORW]
$$
\Biggl(
{{\sin\bigl({1 \over 2}v^{(a)}_j+{i\over 2t_a}s\lambda\delta_{a p}\bigr)}
     \over
{\sin\bigl({1 \over 2}v^{(a)}_j-{i\over 2t_a}s\lambda\delta_{a p}\bigr)}}
\Biggr)^N
   = \prod_{b=1}^r \prod_k
{{\sin\bigl({1\over 2}(v^{(a)}_j-v^{(b)}_k+
i(\alpha_a|\alpha_b)\lambda)\bigr)}
       \over
{\sin\bigl(
{1 \over 2}(v^{(a)}_j-v^{(b)}_k-i(\alpha_a|\alpha_b)\lambda)\bigr)}}.
                                          \eqno{(2.26)}
$$
Here, the integers $1 \le p \le r$ and $s \ge 1$
signify that the transfer matrices $T^{(a)}_m(u)$ act on the
fixed quantum space $W^{(p) \otimes N}_s$
as noted before (2.3).
Note that
$$X_r \neq A_{r \ge 2}, \quad 0 < u < {1 \over 2}
\eqno(2.27{\rm a})$$
as in subsection 2.2.
In addition, we shall restrict ourselves to the case
$$t_p = 1 \eqno(2.27{\rm b})$$
in this subsection for a technical reason and exclusively
study the fusion vertex model
corresponding to the transfer matrix
$T^{(a)}_m(u)$ with $(a,m) = (p,s)$.
Plainly, it is the model whose horizontal and vertical fusion types
are both $W^{(p)}_s$ in the sense of section 3.3 in Part I.
In this case, one may deduce [\rBax] the free
energy from the energy of the corresponding one dimensional quantum
system (cf. eq.(2.19) in [\rKu]) as
$$
F(u) = - \sum_{a=1}^r \sum_j
\log \Biggl(
{{\sin\bigl({1 \over 2}v^{(a)}_j+{i\over 2t_a}s\lambda\delta_{a p}+
i\lambda (u-u^*) \bigr)}
    \over
{\sin\bigl({1 \over 2}v^{(a)}_j-{i\over 2t_a}s\lambda\delta_{a p}+
i\lambda (u-u^*) \bigr)}}
\Biggr).  \eqno{(2.28)}
$$
In the above, $u^*$ signifies the point where
the Hamiltonian limit is to be taken.
In the case of the $sl(r+1)$ models [\rJKMO,\rBRT],
it is given by $1-p$ (see appendix C).
Below, we set $u^*=0$ assuming that its effect
has been absorbed into the spectral parameter $u$
that obeys (2.27a).
Eqs.(2.26,28) are consistent with the analytic Bethe ansatz
in [\rRe] and actually reduces to the known 6-vertex model result for
$X_r = A_1, s=p=1$.
Notice that the $\sin$ function appears
here in contrast to the $\sinh$
function for the ``massless" regime.
In the thermodynamic limit $N \rightarrow \infty$, the solution to
(2.26) is expected to form the pattern
$$
v_j^{(a)}= v^{(a)}_{j,m} +i {{(m+1-2\alpha)\lambda}\over{t_a}},
\quad \alpha = 1,2, \ldots, m
\eqno(2.29)
$$
for some $v^{(a)}_{j,m} \in (-\pi, \pi]$ and $m \in {\bf Z}_{\ge 1}$,
which is called the color $a$ $m$-string with center $v^{(a)}_{j,m}$.
For each color $1 \le a \le r$ and length $m \in {\bf Z}_{\ge 1}$, let
$\rho^{(a)}_m(v)$ and $\sigma^{(a)}_m(v)$ denote the densities of
$m$-strings and $m$-holes with the center $v \in (-\pi, \pi]$, respectively
(cf. [\rYY-\rTS]).
Then eq.(2.26) can be rewritten in terms of these densities
as we did in appendix B of Part I.\par
For a $2\pi$-periodic function $f(v) = f(v + 2\pi)$,
we define its Fourier components as
$$
\hat{f}[n]=\int_{-\pi}^{\pi} f(v) e^{inv} dv\,\,\,
\hbox{ for } \,\, n \in {\bf Z}, \quad
f(v)= {1\over{2\pi}}\sum_{n \in {\bf Z}} \hat{f}[n] e^{-inv}.
\eqno(2.30)
$$
By performing the Fourier transformation with respect to the
centers, (2.26) is rewritten as
$$
\delta_{a p} {\hat{\cal A}}^{s m}_{p a}[n] = \hat{\sigma}^{(a)}_m[n]
+ \sum_{b=1}^{r}\sum_{k\ge 1} {\hat{\cal M}}_{a b}[n]
{\hat{\cal A}}^{m k}_{a b}[n]\hat{\rho}^{(b)}_k[n],
\eqno(2.31)
$$
where
$$\eqalignno{
{\hat {\cal M}}_{a b}[n]& = B_{a b} +
2\delta_{a b}(\cosh{{n\lambda}\over{t_a}}-1),
&(2.32{\rm a})\cr
{\hat {\cal A}}^{m k}_{a b}[n]& =
{\sinh(\hbox{min}({m\over t_a},{k\over t_b})n\lambda)
\over
\sinh{n\lambda\over t_{a b}}}
\exp(-\hbox{max}({m\over t_a}, {k\over t_b})|n|\lambda).
&(2.32{\rm b})\cr}
$$
See (1.2) for the definitions of $B_{a b}$ and $t_{a b}$.
For given string densities, the free energy (2.28) is evaluated as
$$\eqalignno{
{F(u) \over N} &= - \sum_{m \ge 1}\sum_{\alpha = 1}^m \int_{-\pi}^\pi dv
\rho^{(p)}_m(v)
\log \Biggl(
{{\sin\bigl({1 \over 2}v+i{(m+1-2\alpha+s)\lambda \over 2}
+ i\lambda u \bigr)}
    \over
{\sin\bigl({1 \over 2}v+i{(m+1-2\alpha-s)\lambda \over 2}
+ i\lambda u \bigr)}}
\Biggr)\cr
&= \sum_{m \ge 1} \sum_{n \in {\bf Z}}
{e^{-2\lambda n u} {\hat{\cal A}}^{s m}_{p p}[n] \hat{\rho}^{(p)}_m[-n]
\over n},&(2.33)\cr}
$$
for $N \rightarrow \infty$.\par
%
\subskip\noindent
{\it 2.3.2  Excitation energy as a function of hole locations}\par
First we seek the ground state by employing the
string hypothesis:
\vskip0.1cm\pn
\item{(vi)} The ground state is given by the Dirac sea of the color $a$
$st_a$-strings,
\vskip0.1cm\pn
which agrees with many earlier investigations on the special cases
[\rKu-\rBRT,\rBab,\rTakh].
In terms of the density functions, this is equivalent to
$$\eqalignno{
\hat{\sigma}^{(a)}_m[n] &= 0\,\, \hbox{ for all }\,\,
1 \le a \le r\,\,\hbox{ and }\,\, m,&(2.34{\rm a})\cr
\hat{\rho}^{(a)}_m[n] &= 0\,\, \hbox{ unless }\,\, m = st_a.
&(2.34{\rm b})\cr}
$$
Substituting this into (2.31) we find
$$
\hat{\rho}^{(a)}_{st_a}[n]_{gr} = {\hat{\cal Z}}_{a p}[n]
{\hat{\cal A}}^{s s}_{p p}[n],
\eqno(2.35)
$$
where the subscript $gr$ means the ground state and
${\hat{\cal Z}}_{a p}[n]$ is defined by the matrix relation
$$\eqalignno{
\bigl({\hat{\cal Z}}_{a b}[n]\bigr)_{1 \le a,b \le r} &=
\hbox{inverse of }\,
\bigl({\hat{\cal M}}_{a b}[n]
{\hat{\cal A}}^{st_a st_b}_{a b}[n]\bigr)_{1 \le a,b \le r}\cr
&= {e^{s\vert n \vert\lambda}\over \sinh(sn\lambda)}\cdot
\bigl(\hat{{\cal D}}[n]^{-1}\bigr)_{1 \le a,b \le r},
&(2.36{\rm a})\cr
\hat{{\cal D}}[n] &=
\Bigl({{\hat{\cal M}}_{a b}[n]\over \sinh({n\lambda \over t_{a b}})}
\Bigr)_{1 \le a,b \le r}.
&(2.36{\rm b})\cr}
$$
The explicit form of
the inverse matrix $\hat{{\cal D}}[n]^{-1}$ is listed in appendix B
for all the simple Lie algebras $X_r$.
\par
Next, we turn to the excited states
that are obtained by making ``holes" in the sea.
$$
\sigma^{(a)}_{st_a}(v)= {1 \over N} \sum_j \delta(v-\theta^{(a)}_j)
\qquad 1\le a \le r,
\eqno(2.37)
$$
where $\theta^{(a)}_j \in (-\pi,\pi]$ specifies the
location of a hole.
The excitation energy should
be given as a function
$\Delta F(u;\{\theta^{(a)}_j\})$ of the hole locations $\{\theta^{(a)}_j \}$
and the spectral parameter $u$.
To find its explicit form,
solve (2.31) in the presence of (2.37) but still keeping (2.34b).
The result reads
$$
\hat{\rho}^{(a)}_{st_a}[n] = \hat{\rho}^{(a)}_{st_a}[n]_{gr} -
\sum_{b=1}^r\sum_j
{{{\hat{\cal Z}}_{a b}[n]}\over N}e^{in\theta^{(b)}_j},
\eqno(2.38)
$$
giving the deviations of the string densities
from their ground state values.
To be precise, we have neglected here the contributions from
``nearest", ``intermediate" and ``wide" strings [\rBDV], which we
suppose will not affect the final result for the
excitation energy.
This is indeed the case for $X_r = A_1$ as shown in [\rBDV-\rJSu].
By combining (2.32-34), (2.36) and (2.38),
the excitation
energy $\Delta F(u;\{\theta^{(a)}_j\})$ is
expressed as the infinite series
$$
-\sum_{a=1}^r \sum_j\sum_{n \in {\bf Z}}
{\sinh \bigl(n(-2\lambda u + i\theta^{(a)}_j)\bigr)
 \over n \sinh n\lambda}
(\hat{{\cal D}}[n]^{-1})_{p a}.
\eqno(2.39)
$$
This consists of the contributions from the color $a$ holes (2.37)
for each of $1 \le a \le r$.
However, we find it more natural to modify (2.39) slightly
in order to fit the Dynkin diagram symmetry as follows.
(The technical reason for doing this has been stated in the end of
appendix B.)
$$\eqalign{
\Delta &F(u;\{\theta^{(a)}_j\}) \cr
&= -\sum_{a \in \{1,\ldots,r\}/Aut} \sum_j\sum_{n \in {\bf Z}}
{\sinh \bigl(n(-2\lambda u + i\theta^{(a)}_j)\bigr)
 \over n \sinh n\lambda}
\sum_{\tau \in Aut}
(\hat{{\cal D}}[n]^{-1})_{p \,\tau(a)}.\cr}
\eqno(2.40)
$$
Here, $Aut$ stands for the set of Dynkin diagram
automorphisms of $X_r$.
For our case (2.27a), it is trivial, i.e.,
$Aut = \{ id \}$ except for
$X_r = D_r$ and $E_6$ where
$Aut = \{ id, \tau_0 \}$ and the involution
$\tau_0$ is specified by
$$\eqalign{
\tau_0(a) &= \cases{
a & if $1 \le a \le r-2$\cr
r & if $a = r - 1$\cr
r - 1 & if $a = r$\cr}\quad
\hbox{for }\,\, X_r = D_r,\cr
\tau_0(a) &= \cases{
a & if $a = 6$\cr
6 - a & if $1 \le a \le 5$\cr}\quad
\hbox{for }\,\, X_r = E_6.\cr}
\eqno(2.41)
$$
Correspondingly, the range
$\{1,\ldots,r\}/Aut$ of the $a$-sum in (2.40) is just
$\{1, \ldots, r\}$ itself except when
$X_r = D_r$ and $E_6$.
In these cases one can choose, for example,
$\{1,\ldots,r\}/Aut =
\{1,\ldots,r-1\}$ and $\{1,2,3,6\}$
according as
$X_r = D_r$ and $E_6$, respectively.
As is evident form the expression,
(2.40) measures the excitation energy
of the holes located symmetrically with respect to
$Aut$.
It is real for the distributions of holes
invariant under
$\theta^{(a)}_j\rightarrow -\theta^{(a)}_j$.
The $n$-sum in (2.40) is actually convergent in our regime
$0 < u < {1 \over 2}$ (2.27a) owing to (B.11b).
$\Delta F(u;\{\theta^{(a)}_j\})$ can be further evaluated
explicitly in terms of
Jacobi's elliptic function by the formula
($\vert \omega \vert < {g\lambda\over 2}$)
$$
\sum_{n \in {\bf Z}}
{\sinh n\omega\over n \cosh {ng\lambda\over 2}}
= 2\log\Bigl(-i\sqrt{k}\,\hbox{sn}\bigl(
{iK(k)\over \pi}(\omega + {g\lambda\over 2})\bigr)\Bigr),
\eqno(2.42)
$$
where the modulus $k$ of $\hbox{sn}$ is specified by (2.3b,c).
Substituting (B.12) into (2.40) and performing the $n$-sum
by (2.42) we arrive at
$$\eqalignno{
&e^{-\Delta F(u;\{\theta^{(a)}_j\})} \cr
&\quad =
\prod_{a,j,m}\prod_{\epsilon = \pm 1}
\Bigl(-i\sqrt{k}\,\hbox{sn}\bigl(
{2i\lambda K(k)\over \pi}
(-u +{i\theta^{(a)}_j\over 2\lambda} + \epsilon{\beta^{(p,a)}_m\over 2}
+ {g\over 4})\bigr)
\Bigr)^{z^{(p,a)}_m}&(2.43{\rm a})\cr
&\quad = \prod_{a,j,m}
\Bigl(-k\,\hbox{sn}\bigl(
{K(k)\over \pi}\theta^{(a)}_j - i\eta^{(p,a)}_{m,+}\bigr)
\hbox{sn}\bigl(
{K(k)\over \pi}\theta^{(a)}_j - i\eta^{(p,a)}_{m,-}\bigr)
\Bigr)^{z^{(p,a)}_m},
&(2.43{\rm b})\cr
&\eta^{(p,a)}_{m,\pm} =
{2\lambda K(k) \over \pi}
(-u \pm{\beta^{(p,a)}_m\over 2} + {g\over 4}).
&(2.43{\rm c})\cr}
$$
Here $\beta^{(p,a)}_m \ge 0$ and $z^{(p,a)}_m > 0$ are the
constants defined by the expansion (B.12).
Notice that
${1 \over 2} \le \pm {1 \over 2}\beta^{(p,a)}_m + {g \over 4}
\le {g-1 \over 2}$ from (B.13b).
{}From this and $0 < u < {1 \over 2}$ (2.27a),
it follows that
$$
0 < \eta^{(p,a)}_{m,\pm} < {(g-1)\lambda K(k) \over \pi}
= (1 - {1 \over g}) K^\prime(k) < K^\prime(k),
\eqno(2.44)
$$
where (2.3b) has been used.
Thus from (2.43b) and (2.44) we verify that
$e^{-\Delta F(u;\{\theta^{(a)}_j\})}$ is free of
poles in the range $\theta^{(a)}_j \in (-\pi, \pi]$.
\par
%
\subskip\noindent
{\it 2.3.3 Correlation lengths from the excitation energy}\par
Having obtained
the excitation energy $\Delta F(u;\{\theta^{(a)}_j\})$ (2.43),
we can express the correlation function $G(R)$ explicitly as (cf. [\rJKM])
$$
G(R) - G(\infty) \simeq
\int d\{\theta^{(a)}_j \} \sigma(\{\theta^{(a)}_j \})
e^{-\Delta F(u;\{\theta^{(a)}_j \})R},\eqno(2.45)
$$
where $\sigma(\{\theta^{(a)}_j \})$ is the hole distribution function.
As $g(u_1,u_2)$ in (2.25), its explicit form
is not necessary for our purpose of
extracting the correlation length $\xi$.
In order to determine $\xi$,
we consider the simplest case where the only two holes
of the same color $a$ are excited at
$\pm \theta$, for which (2.43) is certainly real
(cf. [\rJKM]).
Substituting the corresponding
$e^{-\Delta F(u;\{\theta, -\theta\})}$ into (2.45), we get the final result,
$$\eqalign{
&G(R) - G(\infty) \cr
&\simeq k^{z R} \int_{-\pi}^\pi d\theta
\sigma(\{\theta, -\theta\}) \prod_m
\vert \hbox{sn}\bigl({K(k)\over \pi}\theta - i\eta^{(p,a)}_{m,+}\bigr)
\hbox{sn}\bigl({K(k)\over \pi}\theta - i\eta^{(p,a)}_{m,-}\bigr)
\vert^{2z^{(p,a)}_m R}.}
\eqno(2.46)
$$
Here $z = 2\sum_m z^{(p,a)}_m$ is a positive constant thanks to (B.13a).
The integral is finite because the integrand is so as we verified
from (2.44).
For a fixed $0 < k < 1$, the decay of (2.46) as
$R \rightarrow \infty$ will then be controlled by the prefactor
$k^{z R}$.
Since $z$ is an order one constant, it will not be concerned
for our definition of the correlation length as mentioned
after (2.3c).
Therefore from (2.46) we obtain the result (2.3).
Notice the similarity of the $u$-dependence in (2.43b) and (2.24).
\par
In the above calculations, we have actually used the condition
$0 < u < {1 \over 2}$ (2.27a)
in (2.33) and to assure the convergence of (2.40).
However, the result (2.46) seems valid in the wider regime
$0 < u < {g \over 2}$ as we commented in the end of
section 2.1.
To support this, we firstly note that
(2.43b) is actually free of
poles when $0 < u < {g \over 2}$ for any
$\theta^{(a)}_j \in (-\pi, \pi]$.
Secondly,
$\Delta F(u;\{\theta^{(a)}_j\}) > 0$ must hold
for the excitation energy and this can be proved
throughout $0 < u < {g \over 2}$ at $\theta^{(a)}_j = 0$ and $\pi$.
Similar ``extension" of the regime has been observed in p155 of [\rBax].
These arguments imply that eqs.(2.25) and (2.46) actually admit
the same parameter range for $u$.\par
\secskip
%
%
%
%
\centerline{\bf{3. Critical RSOS models}}\pn
\subskip\noindent
{\bf 3.1. General remarks}\par
By critical RSOS models we mean those sketched in
section 3 of Part I [\rPaI].
They are specified by the quantum affine algebra
$U_q(X^{(1)}_r)$ and the three integers
$\ell, p$ and $s$ subject to the condition,
$$\ell \ge 1, \quad 1 \le p \le r\,\,\hbox{ and }\,\,
1 \le s \le t_p\ell - 1.\eqno(3.1{\rm a})
$$
The corresponding one is called the level $\ell$
$U_q(X^{(1)}_r)$ RSOS model with fusion type $W^{(p)}_s$.
Throughout this section
we shall reserve the letters $\ell, p$ and $s$ for this meaning
and use the notations
$$\ell_a = t_a \ell\,\,\hbox{ for }\,\,1 \le a \le r,\quad
G = \{\,(a,m) \,\vert 1 \le a \le r, 1 \le m \le \ell_a - 1 \}
\eqno(3.1{\rm b})$$
in accordance with (I3.3).
The well known Andrews-Baxter-Forrester (ABF) model [\rABF]
corresponds to $X_r = A_1$, $(p,s) = (1,1)$ and belongs to
the hierarchy $X_r=A_r$ with general $(p,s)$ [\rJKMO].
The cases
$X_r = B_r, C_r, D_r$ with $(p,s) = (1,1)$ and
$X_r = G_2$ with $(p,s) = (2,1)$ have also been
constructed in [\rJMO] and [\rKS], respectively.
See also [\rKuO,\rWSN].
\par
As shown in section 2 of Part I,
the restricted $T$-system is indeed valid for $X_r = A_r$ RSOS models.
It is the FRs among the row-to-row
transfer matrices $T^{(a)}_m(u)$ with various auxiliary spaces $W^{(a)}_m$
but acting on the common quantum space $W^{(p) \otimes N}_s$
in the corresponding vertex model picture.
Supposing this for all $X_r$, we shall show how
the central charges can be computed from the $T$-system.
The results perfectly agree with
those in [\rKu-\rBRT].
They are also consistent to the 1-point function results
as argued in [\rKu].
\par
There are two regimes to consider corresponding to $\epsilon = +1$ and
$\epsilon = -1$ in [\rKu-\rBRT].
For the ABF model these are the regime I/II boundary and
the regime III/IV boundary, respectively.
The difference of these two regimes lies in which of the two
terms dominates on the rhs of $T$-system (1.1a) in the
thermodynamic limit $N \rightarrow \infty$.
To explain this more concretely,
we again quote (1.1a) (restricted case) in the form
$${\cal T}_0 = {\cal T}_1 + {\cal T}_{-1},\eqno(3.2)$$
as done in (2.5) for the unrestricted case.
In the above, ${\cal T}_i$'s are products of
the transfer matrices $T^{(a)}_m(u)$'s,
which are dependent on the
horizontal system size $N$.
When evaluating (3.2) on the common eigenvector with the largest
eigenvalue of ${\cal T}_0$, we suppose
$$
\vert{\cal T}_{\epsilon}\vert \gg \vert{\cal T}_{-\epsilon}\vert\,\,
\hbox{ as }\,\,N \rightarrow \infty \,\,
\hbox{ in the regime }\,\,\epsilon = \pm 1.
\eqno(3.3)
$$
This is a natural extension from the
$sl(2)$ case [\rKPO,\rKPT].
In particular for $\epsilon = -1$,
ABF's ground state is anti-ferroelectric like [\rABF],
hence (3.3) is also consistent with the
assumption made after (2.5) for the
underlying vertex model.
{}From (3.3), the bulk eigenvalue must be a solution of
${\cal T}_0 = {\cal T}_{\epsilon}$.
On the other hand,
the ratio ${\cal T}_{-\epsilon}/{\cal T}_{\epsilon}$
will measure the finite size correction, which will yield the central charge
in the corresponding CFT [\rBCN,\rAf].
It is through this route that we determine the
central charge starting from our
restricted $T$-system.
The precise formulation will be given below
for each regime.
We note that such a calculation was firstly done
by Kl\"umper and Pearce in [\rKPO,\rKPT] for
$X_r = A_1$.
They also reproduced the scaling dimensions
predicted earlier in [\rDJKMO].
Thus it is our hope to further extend the
treatment of this paper to derive the scaling dimensions for all the
$X_r$ cases.
We leave it as a future problem but
give a formulation which will be of use
for that purpose as far as possible.
We remark that the dilogarithm conjecture in [\rKN] (cf. appendix D)
is an important step toward this direction.
\par
Concerning the case $X_r = A_r$, we have a proof of the $T$-system
in section 2 of Part I and the explicit result on the
asymptotic form of the bulk eigenvalues in appendix C.
Therefore our derivation of the central charge relies only on
the assumption of the ANZC property which will be explained after (3.7).
\par
In the working below, we find it convenient to
renormalize the spectral parameter in the $T$-system
as $T^{(a)}_m(2iu)_{\hbox{new}} = T^{(a)}_m(u)_{\hbox{old}}$.
In this convention, the $T$-system (1.1) (or (3.2)) becomes
$$\eqalign{
{\cal T}^\prime_0 &= {\cal T}^\prime_1 + {\cal T}^\prime_{-1},\cr
{\cal T}^\prime_0 &= T^{(a)}_m(u-{i\over t_a})T^{(a)}_m(u+{i\over t_a}),
\quad
{\cal T}^\prime_1 = T^{(a)}_{m-1}(u)T^{(a)}_{m+1}(u),\cr
{\cal T}^\prime_{-1} &= g^{(a)}_m(u)
\prod_{b=1}^r{\cal T}(a,b,m,{u \over 2i})^{I_{ab}}
\vert_{\forall T^{(a)}_m(x) \rightarrow T^{(a)}_m(2ix)},\cr}\eqno(3.4)$$
with an obvious redefinition of the scalar function $g^{(a)}_m(u)$.
The same assumption as (3.3) will
still be implied for ${\cal T}^\prime_{\pm 1}$.
We recall that the combination
$Y^{(a)}_m(u) = {\cal T}^\prime_{-1}/{\cal T}^\prime_1$ solves the
restricted $Y$-system (IB.6) due to the connection (I3.19).
%
%
\subskip\noindent
{\bf 3.2. Regime $\epsilon$ = +1}\pn
{\it 3.2.1 Integral equation for the finite size correction}\par
Consider the restricted $T$-system (3.4),
where $T^{(a)}_m(u)$ is to be understood
either as a matrix or its eigenvalues.
For $(a,m) \in G$ (3.1b),
put $Y^{(a)}_m(u) = {\cal T}^\prime_{-1}/{\cal T}^\prime_1$.
Thus from (3.3) we see that $\vert Y^{(a)}_m(u) \vert \ll 1$ in the
present regime $\epsilon = +1$.
In view of this we rewrite the $Y$-system (IB.6) slightly to
exclude $Y^{(a)}_m(u)^{-1}$ as
$$
{Y^{(a)}_m(u-{i\over t_a})Y^{(a)}_m(u+{i \over t_a})\over
Y^{(a)}_{m-1}(u)Y^{(a)}_{m+1}(u)} =
{\prod_{b=1}^r\prod_{k=1}^3 F_k(a,m,b;u)^{I_{a b}\delta_{t_ak,\, t_{ab}}}\over
(1 + Y^{(a)}_{m-1}(u))(1 + Y^{(a)}_{m+1}(u))}.
\eqno(3.5)
$$
Note that the rhs only contains the combinations
$1 + Y^{(a^\prime)}_{m^\prime}(u^\prime)$ hence it is close
to 1 for $N$ large.
To separate the bulk and the finite size correction parts we put
$$
Y^{(a)}_m(u) = {YF^{(a)}_m(u) \over YB^{(a)}_m(u)},
\eqno(3.6)
$$
where the bulk part obeys
$$
{YB^{(a)}_m(u-{i\over t_a})YB^{(a)}_m(u+{i \over t_a})\over
YB^{(a)}_{m-1}(u)YB^{(a)}_{m+1}(u)} = 1. \eqno(3.7)
$$
Moreover we assume that
for the largest eigenvalue, the finite size correction part
$YF^{(a)}_m(u)$ and $1 + Y^{(a)}_m(u)$
are Analytic, Non-Zero in some strip on the
complex $u$-plane including the real axis and asymptotically
Constant as $Re(u) \rightarrow \pm \infty$.
This property is referred to as ANZC [\rKPT].
Unfortunately this is not derivable solely from
the $T$-system but we believe it true based on some arguments in the
$X_r = A_1$ case [\rKPT].
The significance of a function $f(u)$ for being ANZC is that it allows
the Fourier transformation after the logarithmic derivative
$$
(\tilde{\log} f)(x) \buildrel\rm def\over = \int_{-\infty}^\infty du
e^{-iux} {\partial \over \partial u}\log f(u).
\eqno(3.8)
$$
Combining (3.5-7) we get
$$
{YF^{(a)}_m(u-{i\over t_a})YF^{(a)}_m(u+{i \over t_a})\over
YF^{(a)}_{m-1}(u)YF^{(a)}_{m+1}(u)} =
{\prod_{b=1}^r\prod_{k=1}^3 F_k(a,m,b;u)^{I_{a b}\delta_{t_ak,\, t_{ab}}}\over
(1 + Y^{(a)}_{m-1}(u))(1 + Y^{(a)}_{m+1}(u))}.
\eqno(3.9)
$$
Now that the both sides consist of ratios of ANZC functions
one can take the $\tilde{\log}$ of them.
Solving the resulting equation for the finite size
correction part, one finds
$$
\tilde{\log} YF^{(a)}_m = \sum_{(b,k) \in G}
\hat{\Psi}^{m k}_{a b}\, \tilde{\log}(1 + Y^{(b)}_k).
\eqno(3.10)
$$
Here $\hat{\Psi}^{a b}_{m k}$ has been defined in (IB.13) and
we have used the fact that the restricted $Y$-system (IB.6)
can be rewritten in the form (IB.25).
By taking the inverse Fourier transformation and integrating over $u$,
(3.10) becomes
$$
\log Y^{(a)}_m = -\log YB^{(a)}_m + \sum_{(b,k) \in G}
\Psi^{m k}_{a b}*\log (1 + Y^{(b)}_k) + \pi i D^{(a)}_m,
\eqno(3.11)
$$
where the last term is the integration constant and
the symbol $*$ is the convolution
as specified in (IB.9).
\vskip0.3cm\pn
{\it 3.2.2 Integral equation in the scaling limit}\par
In (3.11) the dependence on the system size $N$ enters
only through $\log YB^{(a)}_m$.
For $X_r = A_r$, we have determined the explicit asymptotic behavior of this
quantity in appendix C as follows.
$$
\lim
\log YB^{(a)}_m(\pm(u + {\ell \over \pi}\log N))
\equiv 2\delta_{p a}
{\sin{\pi m\over \ell_p}\sin{\pi s\over \ell_p}
\over\sin{\pi \over \ell_p}} e^{-{\pi u\over \ell}} + O({1\over N})
\,\,\hbox{ mod } \,\,2\pi i {\bf Z},
\eqno(3.12)
$$
where the limit $N \rightarrow +\infty$ is taken
under the condition $N \in i_0{\bf Z}$
with $i_0 = 2\ell$.
See (C.13).
Supported by this we assume (3.12) for a general $X_r$ with some integer $i_0$.
This is consistent with (3.7).
Correspondingly we introduce the functions
in the same scaling limit
$$y^{(a)}_{m,\pm}(u) = \lim
Y^{(a)}_m(\pm(u + {\ell \over \pi}\log N)).
\eqno(3.13)
$$
Dropping the $2\pi i {\bf Z}$ part in (3.12) that can be absorbed into
the branch choice, one finds
from (3.11) and (3.13) that $y^{(a)}_{m,\pm}(u)$ satisfies the integral
equation
$$
\log y^{(a)}_{m,\pm} = -2\delta_{p a}
{\sin{\pi m\over \ell_p}\sin{\pi s\over \ell_p}
\over\sin{\pi \over \ell_p}} e^{-{\pi u\over \ell}} + \sum_{(b,k) \in G}
\Psi^{m k}_{a b}*\log (1 + y^{(b)}_{k,\pm}) + \pi i D^{(a)}_m.
\eqno(3.14)
$$
\vskip0.3cm\pn
{\it 3.2.3 Finite size correction to $T^{(a)}_m(u)$ and the
effective central charge}\par
Let us split $T^{(a)}_m(u)$ into the
bulk and the finite size correction parts as
$$
T^{(a)}_m(u) = TB^{(a)}_m(u)TF^{(a)}_m(u).
\eqno(3.15)
$$
On the bulk part we impose
$$
TB^{(a)}_m(u-{i\over t_a})TB^{(a)}_m(u+{i\over t_a}) =
TB^{(a)}_{m-1}(u)TB^{(a)}_{m+1}(u),\eqno(3.16)
$$
which is the same form as (3.7).
{}From the $T$-system (3.4)
and (3.15-16) it follows that
$$
{TF^{(a)}_m(u-{i\over t_a})TF^{(a)}_m(u+{i \over t_a}) \over
TF^{(a)}_{m-1}(u)TF^{(a)}_{m+1}(u)} =
1 + Y^{(a)}_m(u).\eqno(3.17)
$$
Assuming that $TF^{(a)}_m(u)$ is ANZC as well, one can solve this
for it by following the similar procedure to that from (3.9) to (3.11).
The result reads
$$\log TF^{(a)}_m - C^{(a)}_m= \sum_{n=1}^{\ell_a-1}
{\cal A}^{m n}_{a a}*\log(1 + Y^{(a)}_n),
\eqno(3.18)
$$
where $C^{(a)}_m$ is an integration constant and
${\cal A}^{m n}_{a a}$ has been defined in (IB.11) by its
Fourier component.
To see the behavior in the limit $N \rightarrow \infty$,
we rewrite the rhs further as
$$\eqalign{
&\sum_{n=1}^{\ell_a-1}\int_{-{\ell\over \pi}\log N}^\infty
dv {\cal A}^{m n}_{a a}(u+v+{\ell\over \pi}\log N)
\log\bigl(1+Y^{(a)}_n(-v-{\ell\over \pi}\log N)\bigr) \cr
&\,\,+ \sum_{n=1}^{\ell_a-1}\int_{-{\ell\over \pi}\log N}^\infty
dv {\cal A}^{m n}_{a a}(u-v-{\ell\over \pi}\log N)
\log\bigl(1+Y^{(a)}_n(v+{\ell\over \pi}\log N)\bigr). \cr
}\eqno(3.19)
$$
By combining (3.13), (3.19) and
(IB.15), the following scaling behavior
can be derived for $N$ large.
$$\eqalign{
&\log TF^{(a)}_m(u) \cr
&= {\sin{\pi m\over \ell_a} \over N \ell\sin{\pi \over \ell_a}}
\sum_{\kappa = \pm 1}
e^{\kappa\pi u \over \ell}
\sum_{n=1}^{\ell_a-1}\sin{\pi n\over \ell_a}
\int_{-\infty}^\infty
dv e^{-{\pi v \over \ell}}
\log\bigl(1+y^{(a)}_{n,\kappa}(v)\bigr) + C^{(a)}_m + o({1\over N}).\cr}
\eqno(3.20)
$$
We are interested in the non-trivial finite size correction
proportional to $1/N$ for $(a,m) = (p,s)$ in the above.
Following appendix C of [\rKPT], the central charge $c$ and the
scaling dimensions $\Delta_{\pm}$ may be deduced as
$$\eqalign{
\log TF^{(p)}_s(u) &= {\pi c \over 6N}\hbox{cosh}{\pi u \over \ell}
- {2\pi \over N}\Bigl(
(\Delta_+ + \Delta_-)\hbox{cosh}{\pi u \over \ell} +
(\Delta_+ - \Delta_-)\hbox{sinh}{\pi u \over \ell}\Bigr)\cr
&= {\pi \over 12N}e^{\pi u \over \ell}(c-24\Delta_+) +
{\pi \over 12N}e^{-{\pi u \over \ell}}(c-24\Delta_-),\cr
}\eqno(3.21)$$
up to $o(1/N)$ terms.
Here the combinations $c_{{\rm eff},\pm} = c-24\Delta_\pm$
are the effective central charges
for the two chiral halves.
Comparing (3.21) with (3.20) one gets
$$
c_{{\rm eff},\pm} =
{12 \sin{\pi s\over \ell_p} \over \pi\ell \sin{\pi \over \ell_p}}
\sum_{m=1}^{\ell_p-1} \sin{\pi m\over \ell_p}
\int_{-\infty}^\infty du e^{-{\pi u \over \ell}}
\log\bigl(1 + y^{(p)}_{m,\pm}(u)\bigr).
\eqno(3.22)
$$
\vskip0.3cm\pn
{\it 3.2.4 Effective central charge as the Rogers dilogarithm}\par
Let us evaluate the integral in (3.22) by means of
the integral equation in the scaling
limit (3.14).
As it turns out, the result is expressed
in terms of the Rogers dilogarithm.
Consider the identity
$$\eqalign{
&\sum_{(a,m)\in G}\int_{-\infty}^\infty du
\Bigl(\log(1+y^{(a)}_{m,\pm}(u)){\partial\over\partial u}
\log y^{(a)}_{m,\pm}(u) \cr
&\qquad \qquad - \log(y^{(a)}_{m,\pm}(u)){\partial\over\partial u}
\log(1+y^{(a)}_{m,\pm}(u))\Bigr)\cr
&= 2\sum_{(a,m)\in G} \Biggl(
L\Bigl({y^{(a)}_{m,\pm}(\infty)\over 1+y^{(a)}_{m,\pm}(\infty)}\Bigr)
- L\Bigl({y^{(a)}_{m,\pm}(-\infty)\over 1+y^{(a)}_{m,\pm}(-\infty)}\Bigr)
\Biggr),\cr}\eqno(3.23)
$$
which follows directly from the definition (D.2) of the
Rogers dilogarithm.
In the lhs of (3.23), replace the
$\log y^{(a)}_{m,\pm}(u)$ by the rhs of (3.14).
All the terms involving
$\Psi^{n k}_{a b}$ vanishes from the property (IB.14).
Partially integrating the remaining parts, one can derive
$$\eqalign{
&
{2\pi \sin{\pi s\over \ell_p} \over \ell \sin{\pi \over \ell_p}}
\sum_{m=1}^{\ell_p-1} \sin{\pi m\over \ell_p}
\int_{-\infty}^\infty du e^{-{\pi u \over \ell}}
\log\bigl(1 + y^{(p)}_{m,\pm}(u)\bigr)\cr
& = \sum_{(a,m) \in G}\Biggl\lbrack
L\Bigl({y^{(a)}_{m,\pm}(u)\over 1+y^{(a)}_{m,\pm}(u)}\Bigr) -
{\pi i\over 2}D^{(a)}_m
\log\Bigl({1\over 1+y^{(a)}_{m,\pm}(u)}\Bigr)
\Biggr\rbrack^{u = +\infty}_{u = -\infty}.\cr}
\eqno(3.24)
$$
Up to some trivial factor, the lhs of (3.24) is
precisely the combination appearing in (3.22).
Thus we have
$${\pi^2 \over 6}c_{{\rm eff},\pm}
= \sum_{(a,m) \in G}\Biggl\lbrack
L\Bigl({y^{(a)}_{m,\pm}(u)\over 1+y^{(a)}_{m,\pm}(u)}\Bigr) -
{\pi i\over 2}D^{(a)}_m
\log\Bigl({1\over 1+y^{(a)}_{m,\pm}(u)}\Bigr)
\Biggr\rbrack^{u = +\infty}_{u = -\infty}.
\eqno(3.25{\rm a})
$$
The constant $D^{(a)}_m$ here can be determined from (3.14) as
$$
\pi i D^{(a)}_m = \log
\Bigl({y^{(a)}_{m,\pm}(\infty)\over 1+y^{(a)}_{m,\pm}(\infty)}\Bigr) -
\sum_{(b,k) \in G} K^{m k}_{a b}
\log\Bigl({1\over 1+y^{(b)}_{k,\pm}(\infty)}\Bigr),
\eqno(3.25{\rm b})
$$
which must be independent of the $\pm$ suffix.
In deriving (3.25b) we have assumed that $y^{(a)}_{m,\pm}(u)$
approaches its limit rapidly enough when $u \rightarrow \infty$.
Since $\Psi^{m k}_{a b}(u)$ is decaying
for $\vert u \vert$ large,
$\sum_{(b,k) \in G}\Psi^{m k}_{a b}*
\log (1 + y^{(b)}_{k,\pm})$ could then be replaced by
$\sum_{(b,k) \in G}
\hat{\Psi}^{m k}_{a b}(0)\,\log (1 + y^{(b)}_{k,\pm}(\infty))$
for which one can use (IB.13) and (IB.21a).
Eq.(3.25) and (3.14) are the
key ingredients in deriving the
central charge and the scaling dimensions.
\vskip0.3cm\pn
{\it 3.2.5 Central charges}\par
As in [\rKN,\rKNS], (3.25a) may be interpreted as defining a function
that depends on the ways the rhs is analytically continued.
To study the resulting spectrum of $c_{{\rm eff},\pm}$ will be an
interesting problem and it has been argued in appendix D
for some simplified case
(formally without $u = -\infty$ term in (3.25a))
in the light of the dilogarithm conjecture in [\rKN].
Here we concentrate on the simplest case
where $c_{{\rm eff},\pm}$ becomes the central charge itself.
Such a situation is realized when
$\forall y^{(a)}_{m,\pm}(u) > 0$ for $-\infty < u < \infty$ and
$\forall D^{(a)}_m = 0$ as in [\rKPO,\rKPT].
Then all the formulas (3.22-25) may be treated
on an equal footing for both indices $\pm$.
We shall henceforth
suppress it and write $c_{{\rm eff},\pm}$ simply as $c$.
Under the condition $\forall y^{(a)}_m(\infty) > 0$,
the eq.(3.25b) with $\forall D^{(a)}_m = 0$
has the unique solution
$$
0 < {y^{(a)}_m(\infty)\over 1+y^{(a)}_m(\infty)}
= f^{(a)}_m < 1,\eqno(3.26)
$$
where $f^{(a)}_m$ is $f^{(a)}_m(z=0)$ of (D.4).
See (D.5).
As detailed in appendices A and B of Part I and [\rKR,\rKu], this quantity is a
purely Lie algebraic data specified by the algebra
$X_r$ and the level $\ell \in {\bf Z}_{\ge 1}$.
To seek the value of $y^{(a)}_m(-\infty)$,
consider next (3.14) (with $D^{(a)}_m = 0$)
in the limit $u \rightarrow -\infty$.
The negative divergence from the first term on its rhs
will be compensated by the behavior
$$
y^{(p)}_m(u) \buildrel u\rightarrow -\infty \over
\longrightarrow +0
\quad\hbox{ for }\,\, 1 \le m \le \ell_p-1.
\eqno(3.27{\rm a})
$$
Eq.(3.14) is then valid when $u \rightarrow -\infty$ if
the other $y^{(a)}_m(-\infty)$'s obey the reduced equation
$$
0 = \log
\Bigl({y^{(a)}_m(-\infty)\over 1+y^{(a)}_m(-\infty)}\Bigr) -
\sum_{(b,k) \in G, b \ne p} K^{m k}_{a b}
\log\Bigl({1\over 1+y^{(b)}_k(-\infty)}\Bigr)\,\,
\hbox{ for }\,\, a \ne p.
\eqno(3.27{\rm b})
$$
Eq.(3.27b) can be analyzed similarly to section 3.1 of [\rKu].
To do so we visualize the set
$G$ (3.1b) by a tableau whose $a$-th column
$(1 \le a \le r)$ consists of the $\ell_a-1$ rectangles
each having the depth $1/t_a$ and the width 1.
The elementary rectangle at the $a$-th column and the
$m$-th row corresponds to
$y^{(a)}_m(-\infty)$ (or
${y^{(a)}_m(-\infty)\over 1+y^{(a)}_m(-\infty)}$).
We call such a tableau the $G$-tableau for $(X_r,\ell)$.
See fig.1 in [\rKu] for examples.
As depicted there,
hatch the vertical strip
$G_+ = \{ (p,m) \vert 1 \le m \le \ell_p - 1 \}$
in the $G$-tableau to signify (3.27a).
Removing the strip divides the $G$-tableau
into a few unhatched subdomains.
But each such domain is again identical to a $G$-tableau for some
$(X^\prime_{r^\prime},\ell^\prime)$.
Let ${\cal H}_p(X_r,\ell)$ denote the multiple set of the
pairs $(X^\prime_{r^\prime},\ell^\prime)$ so obtained.
For example,
$$\eqalign{
{\cal H}_1(F_4,3) &= \{ (C_3,3) \},\quad\qquad\quad\,\,
{\cal H}_2(F_4,3) = \{ (A_1,3), (A_2,6) \},\cr
{\cal H}_3(F_4,3) &= \{ (A_2,3), (A_1,6) \},\quad
{\cal H}_4(F_4,3) = \{ (B_3,3) \},\cr
}\eqno(3.28{\rm a})
$$
corresponding to (3.7a) in [\rKu].
${\cal H}_p(X_r,\ell)$ has a multiplicity when the $X_r$
Dynkin diagram has a branch, e.g.,
$$\eqalign{
{\cal H}_1(E_6,2) &= {\cal H}_5(E_6,2) = \{ (D_5,2) \},
\qquad\qquad\,{\cal H}_6(E_6,2) = \{ (A_5,2) \},\cr
{\cal H}_2(E_6,2) &= {\cal H}_4(E_6,2) = \{ (A_1,2), (A_4,2) \},\cr
\quad{\cal H}_3(E_6,2) &= \{ (A_1,2), (A_2,2), (A_2,2) \}.\cr
}\eqno(3.28{\rm b})
$$
Returning to (3.27b),
we now see that it decouples into independent
systems of equations
and each of them again takes the form (D.5b) for the pairs
$(X^\prime_{r^\prime},\ell^\prime) \in {\cal H}_p(X_r,\ell)$.
Therefore ${y^{(a)}_m(-\infty)\over 1+y^{(a)}_m(-\infty)}$
contained in the corresponding
$G$-tableau is identified with some
$f^{(a^\prime)}_{m^\prime}$ attached to the
$(X^\prime_{r^\prime},\ell^\prime)$.
In this way the values of $y^{(a)}_m(-\infty)$'s
are specified for all $(a,m) \in G$.
Substituting them and (3.26) into (3.25a) and using
$D^{(a)}_m = 0$ we find
$${\pi^2\over 6}c =
\sum_{(a,m)\in G}L(f^{(a)}_m) -
\sum_{\scriptstyle(X^\prime_{r^\prime},\ell^\prime) \in
{\cal H}_p(X_r,\ell)\atop
\scriptstyle(a,m)\in G \hbox{ for }
(X^\prime_{r^\prime},\ell^\prime)} L(f^{(a)}_m \hbox{ for }
(X^\prime_{r^\prime},\ell^\prime)). \eqno(3.29)
$$
With the aid of the dilogarithm conjecture (D.6), (3.29) is
computed explicitly as
$$
c =
{\ell \hbox{ dim }X_r \over \ell + g} - r -
\sum_{\scriptstyle(X^\prime_{r^\prime},\ell^\prime) \in
{\cal H}_p(X_r,\ell)}
\Bigl({\ell^\prime \hbox{ dim }X^\prime_{r^\prime}
\over \ell^\prime + g^\prime} - r^\prime\Bigr),
\eqno(3.30)
$$
where $g^\prime$ denotes the dual Coxeter number of
$X^\prime_{r^\prime}$.
The result (3.30) completely reproduces the
central charge (3.9) of [\rKu].
It depends on the fusion type $W^{(p)}_s$ only through $p$.
\vskip0.6cm\pn
{\bf 3.3. Regime $\epsilon$ = -1}\par
Here we shall exclusively deal with the
simply laced case $X_r = A_r, D_r$ and $E_{6,7,8}$ to avoid
a technical complexity.
Then the analysis is fairly
parallel to the regime $\epsilon = +1$.
We note however that it is quite straightforward
to write down the integral equation
analogous to (3.11) for any $X_r$.
\par
\vskip0.3cm\pn
{\it 3.3.1 Integral equation for the finite size correction}\par
Throughout section 3.3, we set
$$
Y^{(a)}_m(u) = {T^{(a)}_{m+1}(u)T^{(a)}_{m-1}(u)\over
g^{(a)}_m(u) \prod_{b=1}^r \bigl(T^{(b)}_m(u)\bigr)^{I_{ab}}},\eqno(3.31)
$$
which is the inverse of the definition
$Y^{(a)}_m(u) = {\cal T}^\prime_{-1}/{\cal T}^\prime_1$
in the previous section 3.2.
Then the level $\ell$ restricted $Y$-system (IB.30) is changed into
$$Y^{(a)}_m(u+i)Y^{(a)}_m(u-i) =
{\prod_{k=1}^{\ell-1}\bigl(1 + Y^{(a)}_k(u)\bigr)^{{\bar I}_{mk}}
\over
\prod_{b=1}^r\bigl(1 + Y^{(b)}_m(u)^{-1}\bigr)^{I_{ab}}},
\eqno(3.32)
$$
where $I$ is the incidence matrix (1.2) and
${\bar I}$ denotes the $I$ for $A_{\ell-1}$ as in appendix B.5 of Part I.
The definition (3.31) corresponds to
$Y^{(a)}_m(u) = {\cal T}^\prime_1/{\cal T}^\prime_{-1}$ in the
notation of (3.4) hence
$\vert Y^{(a)}_m(u) \vert \ll 1$
from (3.3) in the regime $\epsilon = -1$ under consideration.
As in (3.5) let us rewrite (3.32) so as to exclude
$Y^{(a)}_m(u)^{-1}$ as
$${Y^{(a)}_m(u+i)Y^{(a)}_m(u-i) \over
\prod_{b=1}^r Y^{(b)}_m(u)^{I_{ab}}}
= {\prod_{k=1}^{\ell-1}\bigl(1 + Y^{(a)}_k(u)\bigr)^{{\bar I}_{mk}}
\over
\prod_{b=1}^r\bigl(1 + Y^{(b)}_m(u)\bigr)^{I_{ab}}}
\quad \hbox{ for }\,\, (a,m) \in G,
\eqno(3.33)
$$
in which the both sides are close to 1 for $N$ large.
We separate the bulk and the finite size correction parts as
$$
Y^{(a)}_m(u) = YB^{(a)}_m(u)YF^{(a)}_m(u),
\eqno(3.34)
$$
where the former satisfies
$${YB^{(a)}_m(u+i)YB^{(a)}_m(u-i) \over
\prod_{b=1}^r YB^{(b)}_m(u)^{I_{ab}}} = 1.
\eqno(3.35)
$$
The $Y^{(a)}_m$'s on the lhs of (3.33) can be
replaced with $YF^{(a)}_m$'s due to (3.34,35).
Assuming that the $YF^{(a)}_m(u)$ and $1+Y^{(a)}_m(u)$ are ANZC,
one can solve the resulting equation by taking the $\tilde{\log}$
on both sides.
This is a similar calculation to (3.8-11) leading to
$$
\log Y^{(a)}_m = \log YB^{(a)}_m + \sum_{(b,k) \in G}
\Phi^{m k}_{a b}*\log (1 + Y^{(b)}_k) + \pi i D^{(a)}_m,
\eqno(3.36)
$$
where $D^{(a)}_m$ is an integration constant.
$\Phi^{m k}_{a b}$ is defined by (IB.31) and has the same property as (IB.14).
The equation (3.36) is the regime $\epsilon = -1$ analogue
of (3.11).
\vskip0.3cm\pn
{\it 3.3.2 Integral equation in the scaling limit}\par
For the bulk part entering (3.36), we assume the following
large $N$ behavior
$$
\lim
\log YB^{(a)}_m(\pm(u + {g \over \pi}\log N))
\equiv -\delta_{s m}
{g \chi^{(a)} \chi^{(p)}\over\sin{\pi \over g}}
e^{-{\pi u\over g}} + O({1\over N})
\,\,\hbox{ mod } \,\,2\pi i {\bf Z},\eqno(3.37)
$$
where the limit $N \rightarrow \infty$ is taken under
$N \in i_0{\bf Z}$ for some $i_0$.
$\chi = (\chi^{(a)})_{1 \le a \le r}$ is the
normalized Perron-Frobenius eigenvector
of the incidence matrix $I_{a b}$
as in (IB.35).
This has been established actually for $X_r = A_r$ in appendix C
where one has $g = r+1$ and
$\chi^{(a)} = \sqrt{2\over r+1}\sin{\pi a \over r+1}$.
Compare (3.37) and (C.12) with $N \in 2g{\bf Z}$.
Eq.(3.37) is also consistent with (3.35)
due to (IB.35a).
Introducing the functions
in the same scaling limit
$$y^{(a)}_{m,\pm}(u) = \lim
Y^{(a)}_m(\pm(u + {g \over \pi}\log N)),
\eqno(3.38)
$$
one derives from (3.36-37) the integral equation
$$
\log y^{(a)}_{m,\pm} = -\delta_{s m}
{g \chi^{(a)} \chi^{(p)}\over\sin{\pi \over g}}
e^{-{\pi u\over g}} + \sum_{(b,k) \in G}
\Phi^{m k}_{a b}*\log (1 + y^{(b)}_{k,\pm}) + \pi i D^{(a)}_m,
\eqno(3.39)
$$
where the $2\pi i{\bf Z}$ part in (3.37) has been dropped
by the same reason as in (3.14).
\vskip0.3cm\pn
{\it 3.3.3 Finite size correction to $T^{(a)}_m(u)$ and the
effective central charge}\par
Let us separate $T^{(a)}_m(u)$ into the
bulk and the finite size correction parts as in (3.15),
where the former is to fulfill the regime
$\epsilon = -1$ analogue of (3.16) as
$$
TB^{(a)}_m(u-i)TB^{(a)}_m(u+i) = g^{(a)}_m(u) \prod_{b=1}^r
TB^{(b)}_m(u)^{I_{ab}}.\eqno(3.40)
$$
In view of (3.15), (3.31) and (3.40), the $T$-system (3.4)
is now equivalent to
$$
{TF^{(a)}_m(u+i)TF^{(a)}_m(u-i) \over
\prod_{b=1}^r TF^{(b)}_m(u)^{I_{ab}}} = 1 + Y^{(a)}_m(u).
\eqno(3.41)
$$
By assuming that $TF^{(a)}_m(u)$ is ANZC, (3.41) can be
transformed into an analogous equation to (3.18) as follows.
$$
\log TF^{(a)}_m - C^{(a)}_m= \sum_{b=1}^r
\,\check{}\,\bigl({\hat {\cal M}}^{-1}\bigr)_{ab}*\log(1 + Y^{(b)}_m),
\eqno(3.42)
$$
where $C^{(a)}_m$ is an integration constant.
The symbol $\,\check{}\,\bigl({\hat {\cal M}}^{-1}\bigr)_{ab}$ denotes the
inverse Fourier transformation (cf.\ (IB.8)) of the $(a,b)$-element
of the matrix ${\hat {\cal M}}^{-1}$ inverse to (IB.10).
Apply the asymptotic form (IB.34) to (3.42).
Then the large $N$ behavior of $\log TF^{(a)}_m$ can
be deduced in a similar manner to (3.18-20) as
$$\eqalign{
&\log TF^{(a)}_m(u) \cr
&= {\chi^{(a)} \over 2N\sin{\pi \over g}}
\sum_{\kappa = \pm 1}
e^{\kappa\pi u \over g}
\sum_{b=1}^r\chi^{(b)}
\int_{-\infty}^\infty
dv e^{-{\pi v \over g}}
\log\bigl(1+y^{(b)}_{m,\kappa}(v)\bigr) + C^{(a)}_m + o({1\over N}).\cr}
\eqno(3.43)
$$
This is the finite size correction proportional to $1/N$ from which
the central charge $c$ and the scaling dimensions
$\Delta_{\pm}$ are to be extracted.
Guided by appendix B of [\rKPT],
we postulate the equation (3.21) with
$\ell$ replaced by $g$ in the present regime $\epsilon = -1$.
Then the effective central charges
$c_{{\rm eff},\pm} = c-24\Delta_\pm$
for the two chiral halves are given by
$$
c_{{\rm eff},\pm} =
{6 \chi^{(p)} \over \pi \sin{\pi \over g}}
\sum_{a=1}^r \chi^{(a)}
\int_{-\infty}^\infty du e^{-{\pi u \over g}}
\log\bigl(1 + y^{(a)}_{s,\pm}(u)\bigr).
\eqno(3.44)
$$
\vskip0.3cm\pn
{\it 3.3.4 Effective central charge as the Rogers dilogarithm}\par
The integral in (3.44)  can be evaluated in a similar way to
section 3.2.4 by considering the identity (3.23).
Replacing $\log y^{(a)}_{m,\pm}(u)$ there by the rhs of (3.39) one can show
$$\eqalign{
&
{\pi \chi^{(p)} \over \sin{\pi \over g}}
\sum_{a=1}^r \chi^{(a)}
\int_{-\infty}^\infty du e^{-{\pi u \over g}}
\log\bigl(1 + y^{(a)}_{s,\pm}(u)\bigr)\cr
& = -\sum_{(a,m) \in G}\Biggl\lbrack
L\Bigl({1\over 1+y^{(a)}_{m,\pm}(u)}\Bigr) +
{\pi i\over 2}D^{(a)}_m
\log\Bigl({1\over 1+y^{(a)}_{m,\pm}(u)}\Bigr)
\Biggr\rbrack^{u = +\infty}_{u = -\infty},\cr}
\eqno(3.45)
$$
which is the regime $\epsilon = -1$ analogue of (3.24).
Combining (3.44) and (3.45) we find the
expression of $c_{{\rm eff},\pm}$ in terms of the Rogers dilogarithm
$${\pi^2 \over 6}c_{{\rm eff},\pm}
= -\sum_{(a,m) \in G}\Biggl\lbrack
L\Bigl({1\over 1+y^{(a)}_{m,\pm}(u)}\Bigr) +
{\pi i\over 2}D^{(a)}_m
\log\Bigl({1\over 1+y^{(a)}_{m,\pm}(u)}\Bigr)
\Biggr\rbrack^{u = +\infty}_{u = -\infty}.
\eqno(3.46{\rm a})
$$
The constant $D^{(a)}_m$ is specified from the
$u \rightarrow +\infty$ limit of (3.39) as
$$\eqalign{
&\pi i D^{(a)}_m \cr
&= -\sum_{(b,k) \in G}
\bigl(C^{-1}\bigr)_{ab} {\bar C}_{m k}\Biggl(
\log\Bigl({1\over 1+y^{(b)}_{k,\pm}(\infty)}\Bigr) -
\sum_{(c,j) \in G} K^{k j}_{b c}
\log\Bigl({y^{(c)}_{j,\pm}(\infty)\over 1+y^{(c)}_{j,\pm}(\infty)}\Bigr)
\Biggr)\cr}
\eqno(3.46{\rm b})
$$
and this should be independent of the suffix $\pm$.
In (3.46b) we have followed a similar
procedure to the derivation of (3.25b) and used
(IB.32c,d) for simply laced $X_r$.
Eq.(3.46) and (3.39) are our basic ingredients
in studying the central charge and the scaling dimensions
in the regime $\epsilon = -1$.
\vskip0.3cm\pn
{\it 3.3.5 Central charges}\par
As in the regime $\epsilon = +1$ we shall hereafter
restrict our analysis
of (3.46) to the simplest case
$\forall y^{(a)}_{m,\pm}(u) > 0$ for $-\infty < u < \infty$ and
$\forall D^{(a)}_m = 0$,
which corresponds to the
central charge $c_{{\rm eff},\pm} = c$.
We suppress the $\pm$ indices and the
identical argument is to be implied for both choices.
Investigations of the full aspects of (3.46) and (3.39)
will be an interesting future problem relevant to the
scaling dimensions.
\par
The constraints $\forall y^{(a)}_m(u) > 0$ and
$\forall D^{(a)}_m = 0$ on (3.46b)
uniquely determine the limit $y^{(a)}_m(\infty)$ as
$$
0 < {1 \over 1 + y^{(a)}_m(\infty)} = f^{(a)}_m < 1,
\eqno(3.47)
$$
where $f^{(a)}_m$ is again $f^{(a)}_m(z=0)$ of (D.4).
On the other hand, the negative divergence from
the $u \rightarrow -\infty$ limit of the
$e^{-{\pi u \over g}}$-term
in (3.39) compels
$$
y^{(a)}_s(u) \buildrel u\rightarrow -\infty \over
\longrightarrow +0
\quad\hbox{ for }\,\, 1 \le a \le r.
\eqno(3.48{\rm a})
$$
Then (3.39) (with $D^{(a)}_m = 0$) holds for $u \rightarrow -\infty$ if
the other $y^{(a)}_m(-\infty)$'s satisfy
$$
0 = \log
\Bigl({1\over 1+y^{(a)}_m(-\infty)}\Bigr) -
\sum_{(b,k) \in G, k \ne s} K^{m k}_{a b}
\log\Bigl({y^{(b)}_k(-\infty)\over 1+y^{(b)}_k(-\infty)}\Bigr)\,\,
\hbox{ for }\,\, m \ne s.
\eqno(3.48{\rm b})
$$
which can be analyzed similarly to section 3.2.5
with the $G$-tableau for the pair $(X_r, \ell)$.
For the simply laced algebras $X_r$, it has the simple
$(\ell-1) \times r$ rectangular shape.
This time we hatch the horizontal strip
$G_{-1} = \{ (a,s) \vert 1 \le a \le r \}$ in it
to depict (3.48a).
Removal of it splits the $G$-tableau into
those corresponding to $(X_r, s)$ and $(X_r,\ell-s)$.
Accordingly, (3.48b) decouples into the independent
systems (D.5b) for these pairs within
which ${1\over 1+y^{(a)}_m(-\infty)}$ is
determined as in (3.47).
Thus (3.46a) with $D^{(a)}_m = 0$ yields
$$\eqalign{
{\pi^2\over 6}c = &-\sum_{(a,m)\in G}L(f^{(a)}_m) + rL(1)
+ \sum_{(a,m) \in G \,\hbox{ for }\, (X_r,s)}
L(f^{(a)}_m \,\hbox{ for }\, (X_r,s))\cr
&+ \sum_{(a,m) \in G \,\hbox{ for }\, (X_r,\ell-s)}
L(f^{(a)}_m \,\hbox{ for }\, (X_r,\ell-s)).\cr}
\eqno(3.49)
$$
By using the dilogarithm conjecture (D.6) and
$L(1) = {\pi^2\over 6}$, this is evaluated explicitly as
$$
c = {s \hbox{ dim }X_r \over s + g} +
{(\ell - s) \hbox{ dim }X_r \over \ell - s + g}
- {\ell \hbox{ dim }X_r \over \ell + g},
\eqno(3.50)
$$
which corresponds to the coset pair [\rGKO]
$$\eqalign{
&X^{(1)}_r \oplus X^{(1)}_r \supset X^{(1)}_r\cr
\hbox{levels}\,\,&\,\, s\qquad \ell - s\qquad \ell.\cr
}\eqno(3.51)
$$
The result (3.50) agrees with that in [\rKu-\rBRT].
It depends on the fusion type $W^{(p)}_s$ only through $s$.
\secskip
%
%
%
\centerline{\bf 4. Summary and Discussions}\subskip

In this paper, we have shown two applications of our
$T$-system proposed in Part I [\rPaI].
In section 2, it has been used to determine the
correlation length of the massive vertex model
in AF regime associated with any $X_r \neq A_{r \ge 2}$.
By assuming the dominance
$\vert {\cal T}_1 \vert \ll \vert {\cal T}_{-1}\vert$
in (2.5) in the thermodynamic limit,
the problem essentially reduces to finding the
periodicity of the bulk $T$-system (2.6).
We have proved that it is just given by the dual Coxeter number
and thereby obtained the result (2.3).
It generalizes
the earlier ones [\rBax,\rKSZ,\rJKM] for the $sl(2)$ case.
In section 3, we have used the $T$-system to derive the central charges
of the level $\ell$ critical $U_q(X^{(1)}_r)$ RSOS model
with fusion type $W^{(p)}_s$.
This is an extension of the approach in [\rKPO,\rKPT]
for the $sl(2)$ case.
Supposing the ANZC property,
we have converted the $T$-system into the integral equations
(3.14) and (3.39).
Then the effective
central charges $c_{{\rm eff}, \pm}$ are extracted from the
finite size corrections and expressed by the Rogers dilogarithm as
in (3.25) and (3.46), respectively.
Full investigations of these equations will yield all the
scaling dimensions in the underlying CFT.
We have actually restricted our analysis to the
simplest situation where $c_{{\rm eff}, \pm}$
becomes the central charge itself.
Then it has been evaluated explicitly in (3.30) and (3.50)
by means of the dilogarithm conjecture [\rKi,\rKu].
They perfectly reproduce the results in [\rKu-\rBRT].
\par
The central charge calculation in section 3 exhibits
a strong resemblance to the
TBA analyses in [\rKu-\rBRT].
In fact, the both approaches should be equivalent
as the means of central charge calculations
at least spiritually [\rBCN,\rAf].
The TBA treats
the infinite length one-dimensional quantum system and seeks
its specific heat,
which vanishes linearly with
the temperature $\beta^{-1} \rightarrow 0$.
On the other hand,
the FR approach in this paper
effectively deals with the corresponding two dimensional classical system
on an infinite strip of width $N$ and evaluates
the $1/N$ correction to the ground state energy.
So they are essentially the same thing under the
identification $\beta = N \rightarrow \infty$.
Our analysis not only verifies this natural
coincidence but goes beyond.
It manifests that there underlies
a common mathematical structure
between the two approaches, TBA and FR.
The basic ingredients in the two methods
are the $Y$ and the $T$-systems, repspectively and they
are intriguingly connected through (I3.19).
The connection makes
the integral equations (3.14) and (3.39)
possess an almost identical structure to the TBA equation (IB.4),
hence the analyses thereafter essentially the same in the two methods.
It therefore appears ever fundamental to unveil
the full content of the key relation (I3.19)
between the $Y$ and the $T$-systems.
In the technical aspects,
the TBA relies crucially on the string
hypothesis, while the FR needs the ANZC property instead.
These assumptions are yet to be justified in general.
\par
There are several future problems that stem from the
present paper.
Firstly, one should be able to analyze the
full aspects of the effective central charges
(3.25) and (3.46).
As shown in [\rKPT] for the $sl(2)$ case,
it will yield the scaling dimensions
as well as the central charge.
Appendix D is a study in this direction,
leading to a generalized dilogarithm conjecture [\rKN].
One may even stumble to reproduce
the character itself for the underlying CFT
in the spirit of [\rKNS].
Secondly, our $T$-system is expected to hold even in
off-critical RSOS models.
Thus one might hope to derive the
correlation lengths and the interfacial tensions
in those models by extending the
analysis in [\rBP].

\sect{Acknowledgements}
\subskip

This work is supported in part by JSPS fellowship,
NSF grants PHY-87-14654, PHY-89-57162 and Packard fellowship.

\subskip
\vskip0.3cm
%
%
%
%

\centerline{\bf{Appendix A. Periodicity of the bulk $T$-system}}
\subskip\par
Let us continue the proof of the proposition in section
2.2.1.\pn
$X_r=C_r$:
The relevant bulk $T$-system is
$$\eqalignno{
&T^{(a)}_{2m}(u-\qtr) T^{(a)}_{2m}(u+\qtr) =
T^{(a+1)}_{2m}(u) T^{(a-1)}_{2m}(u)\quad\hbox{ for }
\,\, 1 \le a \le r-2,    &({\rm A}.1{\rm a})\cr
&T^{(r-1)}_{2m}(u-\qtr) T^{(r-1)}_{2m}(u+\qtr) = T^{(r)}_m(u-\qtr)
T^{(r)}_{m}(u+\qtr) T^{(r-2)}_{2m}(u), &({\rm A}.1{\rm b})\cr
&T^{(r)}_{m}(u-\hlf)T^{(r)}_{m}(u+\hlf) = T^{(r-1)}_{2m}(u), &({\rm A}.1{\rm
c})
}$$
where $T^{(0)}_{2m}(u) = 1$.
{}From (A.1a) we have
$$
T^{(a)}_{2m}(u)=\prod_{j=1}^{a}T^{(1)}_{2m}(u+{j\over 2}-{{a+1}\over 4})
\eqno{({\rm A}.2)}
$$
for $1 \le a \le r-1$.
Applying it to (A.1b,c) one gets
$$\eqalignno{
T^{(r)}_m(u-\qtr)T^{(r)}_m(u+\qtr)
&=\prod_{j=1}^{r}T^{(1)}_{2m}(u+{{j-1}\over 2}-{{r-1}\over 4}) ,
&({\rm A}.3{\rm a})\cr
T^{(r)}_m(u-\hlf)T^{(r)}_m(u+\hlf)
&= \prod_{j=1}^{r-1}T^{(1)}_{2m}(u+{j\over 2}-{r\over 4}),
&({\rm A}.3{\rm b})\cr}
$$
from which
$T^{(1)}_{2m}(u+r/4) = T^{(r)}_m(u)/T^{(r)}_m(u-1/2)$ follows.
Substituting this back into the rhs of (A.3b) we find
$T^{(r)}_m(u)T^{(r)}_m(u+{{r+1}\over 2}) = 1$.
{}From this and (A.2), (2.8) is easily seen for all $1 \le a \le r$.
\vskip0.2cm\pn
$X_r=D_r$:
The bulk $T$-system is
$$\eqalignno{
&T^{(a)}_m(u-\hlf) T^{(a)}_m(u+\hlf)
    = T^{(a+1)}_m(u) T^{(a-1)}_m(u) \quad \hbox{ for }
\,\,1 \le a\le r-3,     &({\rm A}.4{\rm a})\cr
&T^{(r-2)}_m(u-\hlf) T^{(r-2)}_m(u+\hlf)
   = T^{(r-3)}_m(u) T^{(r-1)}_m(u)T^{(r)}_m(u),   &({\rm A}.4{\rm b})\cr
&T^{(r-1)}_m(u-\hlf) T^{(r-1)}_m(u+\hlf)
   = T^{(r-2)}_m(u), &({\rm A}.4{\rm c})\cr
&T^{(r)}_m(u-\hlf) T^{(r)}_m(u+\hlf)
    = T^{(r-2)}_m(u), &({\rm A}.4{\rm d}) \cr
}$$
where $T^{(0)}_m(u)=1$.
By the same argument as the $X_r=A_r$ case,
(2.11) holds for $1 \le a \le r-2$.
Substituting it into (A.4b), we have
$$
T^{(r-1)}_m(u) T^{(r)}_m(u) =
  \prod_{j=1}^{r-1} T^{(1)}_m(u+j-{r \over 2}).
\eqno({\rm A}.5)
$$
Using this in the product of (A.4c) and (A.4d), we deduce
$T^{(1)}_m(u)T^{(1)}_m(u+r-1)=1$,
from which (2.8) can be shown for all the $T^{(a)}_m(u)$'s.
\vskip0.2cm
\pn
$X_r=E_r, r=6,7,8$:
The bulk $T$-system is
$$T^{(a)}_m(u-{1\over 2})T^{(a)}_m(u+{1 \over 2}) =
\prod_{b=1}^r T^{(b)}_m(u)^{I_{a b}}
\quad \hbox{ for }\, 1 \le a \le r.\eqno({\rm A}.6)
$$
When $r=6$, all the $T^{(a)}_m(u)$'s are
expressible via $T^{(1)}_m(u)$ by combining (A.6).
In addition we have
one consistency condition
$$
T^{(1)}_m(u+3)= T^{(1)}_m(u) T^{(1)}_m(u+1) T^{(1)}_m(u+5)
                T^{(1)}_m(u+6). \eqno({\rm A}.7)
$$
Rewriting $T^{(1)}_m(u+5)$ on the rhs by the same equation, we have
$$
T^{(1)}_m(u) T^{(1)}_m(u+1)  T^{(1)}_m(u+2) =
{1 \over { T^{(1)}_m(u+6) T^{(1)}_m(u+7)  T^{(1)}_m(u+8)}},
\eqno({\rm A}.8)
$$
which leads to
$T^{(1)}_m(u) T^{(1)}_m(u+1) \cdots  T^{(1)}_m(u+11) =1$.
Therefore we obtain (2.7) and (2.8) for $a = 1$.
All the other cases $a \ne 1$ are assured by this. \par
When $r=7$, from (A.6)
all the $T^{(a)}_m(u)$'s are
expressible in terms of $T^{(6)}_m(u)$ which must obey
the consistency condition
$$
T^{(6)}_m(u+3)T^{(6)}_m(u+4) =
T^{(6)}_m(u)T^{(6)}_m(u+1)T^{(6)}_m(u+6) T^{(6)}_m(u+7).
\eqno({\rm A}.9)
$$
One can show $T^{(6)}_m(u)T^{(6)}_m(u+9) = 1$ from this, hence (2.8) is valid.
\par
When $r=8$, from (A.6)
all the $T^{(a)}_m(u)$'s are expressible by $T^{(1)}_m(u)$ which
must satisfy the consistency condition
$$
T^{(1)}_m(u-1) T^{(1)}_m(u)T^{(1)}_m(u+1) =
T^{(1)}_m(u-4) T^{(1)}_m(u-3) T^{(1)}_m(u+3)T^{(1)}_m(u+4).
\eqno({\rm A}.10)
$$
{}From this one can deduce
$$
T^{(1)}_m(u) T^{(1)}_m(u+1) T^{(1)}_m(u+2) =
{1 \over {T^{(1)}_m(u+15) T^{(1)}_m(u+16) T^{(1)}_m(u+17)}},
\eqno({\rm A}.11)
$$
which leads to
$T^{(1)}_m(u) T^{(1)}_m(u+1) \cdots  T^{(1)}_m(u+29)=1$,
proving (2.7) and (2.8).
\vskip0.2cm\pn
$X_r=F_4$:
The relevant bulk $T$-system is
$$\eqalignno{
T^{(1)}_m(u-\hlf)T^{(1)}_m(u+\hlf)
            &= T^{(2)}_{m}(u),   &({\rm A}.12{\rm a})\cr
T^{(2)}_m(u-\hlf)T^{(2)}_m(u+\hlf)
            &= T^{(1)}_m(u) T^{(3)}_{2m}(u),
                            &({\rm A}.12{\rm b})\cr
T^{(3)}_{2m}(u-\qtr)T^{(3)}_{2m}(u+\qtr)
           &= T^{(2)}_m(u-{1\over 4})T^{(2)}_m(u+{1\over 4}) T^{(4)}_{2m}(u),
                            &({\rm A}.12{\rm c})\cr
T^{(4)}_{2m}(u-\qtr)T^{(4)}_{2m}(u+\qtr)
           &= T^{(3)}_{2m}(u).    &({\rm A}.12{\rm d})
}$$
All the $T^{(a)}_m(u)$'s are expressible by $T^{(1)}_m(u)$ which
must satisfy the consistency condition
$$
T^{(1)}_m(u)=T^{(1)}_m(u-{3\over 2})T^{(1)}_m(u+{3\over 2}).
\eqno({\rm A}.13)
$$
This leads to
$T^{(1)}_m(u)T^{(1)}_m(u+{9\over 2})=1$, proving (2.8).
\vskip0.2cm\pn
$X_r=G_2$:
The relevant $T$-system is
$$\eqalignno{
T^{(1)}_m(u-\hlf)T^{(1)}_m(u+\hlf)
           &= T^{(2)}_{3m}(u),   &({\rm A}.14{\rm a})\cr
T^{(2)}_{3m}(u-{1\over 6})T^{(2)}_{3m}(u+{1\over 6}) &=
T^{(1)}_m(u-{1\over 3})T^{(1)}_m(u) T^{(1)}_m(u+{1\over 3}).
&({\rm A}.14{\rm b})\cr }$$
Substituting the first equation into the second we get
$$
T^{(1)}_m(u)=T^{(1)}_m(u-{2\over 3} )T^{(1)}_m(u+{2\over 3}).
\eqno({\rm A}.15)
$$
{}From this it follows that
$T^{(1)}_m(u) T^{(1)}_m(u+2)=1$, proving (2.8).
\secskip
%
%
%

\centerline{
{\bf Appendix B. Inverse of the matrix $\hat{{\cal D}}[n]$}}
\subskip\par
Let $\hat{{\cal D}} = \hat{{\cal D}}[n]$ be
the symmetric matrix defined by (2.32a) and (2.36b).
Below we list the explicit form of the matrix elements of its
inverse
$(\hat{{\cal D}}^{-1})_{a b}$
using the notation
$$
s(i_1,\ldots,i_\alpha) = \prod_{k=1}^\alpha
\sinh(i_k n\lambda),\quad
c(i_1,\ldots,i_\alpha) = \prod_{k=1}^\alpha
\cosh(i_k n\lambda).
\eqno({\rm B}.1)
$$
We shall omit some matrix elements in view of the
symmetry $(\hat{{\cal D}}^{-1})_{a b} = (\hat{{\cal D}}^{-1})_{b a}$.\pn
$X_r = A_r:$
$$(\hat{{\cal D}}^{-1})_{a b} =
{s(\hbox{min}(a,b),r+1-\hbox{max}(a,b))\over s(r+1)},
\eqno({\rm B}.2)
$$
where only the case $r=1$ is relevant to our discussion due to our
assumption $t_p=1$.
\pn
$X_r = B_r:$
$$\eqalign{
(\hat{{\cal D}}^{-1})_{a b} &=
{s(\hbox{min}(a,b))c(r-{1\over 2}-\hbox{max}(a,b))\over c(r-{1\over 2})}
\quad \hbox{for }\,\, 1 \le a,b \le r-1,\cr
(\hat{{\cal D}}^{-1})_{r a} &=
{s(a)\over 2c(r-{1\over 2})}
\quad \hbox{for }\,\, 1 \le a \le r-1,\cr
(\hat{{\cal D}}^{-1})_{r r} &=
{s(r) \over 4c({1\over 2},r-{1\over 2})}.\cr}\eqno({\rm B}.3)
$$
\noindent
$X_r = C_r:$
$$
(\hat{{\cal D}}^{-1})_{a b} =
{s({1\over 2}\hbox{min}(a,b))c({1\over 2}(r+1-\hbox{max}(a,b)))\over
c({1\over 2}(r+1))}.
\eqno({\rm B}.4)
$$
\noindent
$X_r = D_r:$
$$\eqalign{
(\hat{{\cal D}}^{-1})_{a b} &=
{s(\hbox{min}(a,b))c(r-1-\hbox{max}(a,b))\over c(r-1)}
\quad \hbox{for }\,\, 1 \le a,b \le r-2,\cr
(\hat{{\cal D}}^{-1})_{r a} &=
(\hat{{\cal D}}^{-1})_{r-1 a} =
{s(a)\over 2c(r-1)}
\quad \hbox{for }\,\, 1 \le a \le r-2,\cr
(\hat{{\cal D}}^{-1})_{r r-1} &=
{s(r-2)\over 4c(1,r-1)},\cr
(\hat{{\cal D}}^{-1})_{r-1 \, r-1} &= \hat{{\cal L}}_{r r} =
{s(r) \over 4c(1,r-1)}.\cr}\eqno({\rm B}.5)
$$
For the remaining exceptional algebras we shall only present
the upper half of the $\hat{\cal D}^{-1}$.
\noindent
$X_r = E_6:$
$$
\hat{{\cal D}}^{-1} = {1 \over c(6)}\pmatrix{
{s(1,8)\over 2s(3)} & {s(1,5)c(2)\over s(3)} &
{1\over 2}s(4)      & {s(2,4)\over 2s(3)} &
{s(1,4)\over 2s(3)} & s(1)c(2)\cr
                   &{s(4,5)\over 2s(3)}&
s(4)c(1)           &{s(2,4)c(1)\over s(3)}&
{s(2,4)\over 2s(3)}&{1 \over 2}s(4)\cr
&&2s(3)c(1,2)& s(4)c(1) &{1 \over 2}s(4)&
s(3)c(2)\cr
&&&{s(4,5)\over 2s(3)}&{s(1,5)c(2)\over s(3)}&
{1 \over 2}s(4)\cr
&&&&{s(1,8)\over 2 s(3)}&s(1)c(2)\cr
&&&&&{s(4)c(3)\over 2c(1)}\cr}.\eqno({\rm B}.62)
$$
\noindent
$X_r = E_7:$
For a typographical reason we split
$\hat{{\cal D}}^{-1}$ into two pieces.
$$
(\hat{{\cal D}}^{-1})_{1 \le a \le 7,\, 1 \le b \le 4} = {1 \over c(9)}
\pmatrix{
2s(1)c(3,5) & {s(6)c(3)\over 2c(1)} & s(4)c(3) & {1\over 2}s(6)\cr
&s(6)c(3)&2s(4)c(1,3)&s(6)c(1)\cr
&&2s(6)c(1,2)&{s(3,6)\over 2s(1)}\cr
&&&{s(5,6)\over 2s(2)}\cr
&&&\cr},\eqno({\rm B}.7{\rm a})
$$
$$
(\hat{{\cal D}}^{-1})_{1 \le a \le 7,\, 5 \le b \le 7} = {1 \over c(9)}
\pmatrix{
s(2)c(3)& s(1)c(3) & 2s(1)c(2,3)\cr
2s(2)c(1,3)&s(2)c(3)&s(4)c(3)\cr
s(6)c(1)&{1\over 2}s(6)&s(6)c(2)\cr
s(5)c(3)&{s(5)c(3)\over 2c(1)}&{s(3,6)\over 2s(2)}\cr
2s(2)c(3,4)&2s(1)c(3,4)&{1\over 2}s(6)\cr
&{s(12)\over 8c(1,2)}&{s(6)\over 4c(1)}\cr
&&{s(7)c(3)\over 2c(1)}\cr}.\eqno({\rm B}.7{\rm b})
$$
\noindent
$X_r = E_8:$
For a typographical reason we split
$\hat{{\cal D}}^{-1}$ into two pieces.
$$(\hat{{\cal D}}^{-1})_{1 \le a \le 8,\, 1 \le b \le 4} = {1\over c(15)}
\pmatrix{
2s(1)c(5,9)&{s(12)c(5)\over 4c(1,2)}&4s(1)c(3,4,5)&
{s(10)c(3)\over 2c(1)}\cr
&{s(12)c(5)\over 2c(2)}&4s(2)c(3,4,5)&s(10)c(3)\cr
&&2s(6)c(4,5)&{s(6,10)\over 2s(2)}\cr
&&&2s(10)c(2,3)\cr
&&&\cr},\eqno({\rm B}.8{\rm a})
$$
$$\eqalign{
&(\hat{{\cal D}}^{-1})_{1 \le a \le 8,\, 5 \le b \le 8} \cr
&= {1\over
c(15)} \pmatrix{
s(6)c(5)&
2s(2)c(3,5)&2s(1)c(3,5)&{s(6)c(5)\over 2c(1)}\cr
2s(6)c(1,5)&
4s(2)c(1,3,5)&2s(2)c(3,5)&s(6)c(5)\cr
{s(3,6)c(5)\over s(1)}&
2s(6)c(1,5)&s(6)c(5)&{s(3,6)c(5)\over s(2)}\cr
4s(6)c(1,2,5)&4s(4)c(1,3,5)&2s(4)c(3,5)&2s(6)c(2,5)\cr
{s(6,10)\over 2s(1)}&2s(10)c(1,3)&s(10)c(3)&{s(6,10)\over 2s(2)}\cr
&2s(7)c(3,5)&{s(7)c(3,5)\over c(1)}&s(10)c(3)\cr
&&4s(1)c(3,5,6)&{s(10)c(3)\over 2c(1)}\cr
&&&{s(8)c(3,5)\over c(1)}\cr}\cr}.\eqno({\rm B}.8{\rm b})
$$
\noindent
$X_r = F_4:$
$$
\hat{{\cal D}}^{-1}= {1 \over c({9\over 2})}\pmatrix{
2s(1)c({3\over 2},2)&  s(3)c({1\over 2})  &
2s(1)c({1\over 2},{3\over 2}) &s(1)c({3\over 2}) \cr
           &  2s(3)c({1\over 2},1) & 2s(2)c({1\over 2},{3\over 2}) &
 s(2)c({3\over 2}) \cr
           &              & s(3)c({3\over 2}) &
{s(3)c({3\over 2})\over 2c({1\over 2})}\cr
           &             &   &2s({1\over 2})c({3\over 2},{5\over 2})\cr}.
\eqno({\rm B}.9)
$$
\noindent
$X_r = G_2:$
$$
\hat{{\cal D}}^{-1} = {1 \over c(2)}\pmatrix{
2s(1)c({1\over 3},{2\over 3})&  s(1)c({2\over 3})  \cr
           &  2s({1\over 3})c({2\over 3},1)  \cr}.
\eqno({\rm B}.10)
$$
In general the matrix $\hat{{\cal D}}[n]^{-1}$ has the properties
$$\eqalignno{
\hat{{\cal D}}[n]^{-1} &= - \hat{{\cal D}}[-n]^{-1},
&({\rm B}.11{\rm a})\cr
(\hat{{\cal D}}[n]^{-1})_{p a} &=
O(e^{\gamma \vert n \vert \lambda})\quad
\hbox{for some } \quad \gamma \le 0 \,\, \hbox{ as } \,\,
\vert n \vert \rightarrow \infty.
&({\rm B}.11{\rm b})\cr}
$$
When $t_p = 1$, all the combinations appearing in
(2.40) can be expanded into a
finite sum of the form
$$
{\sum_{\tau \in Aut}
(\hat{{\cal D}}[n]^{-1})_{p\, \tau(a)} \over
\sinh n\lambda}  =
\sum_m z^{(p,a)}_m \, {\cosh \beta^{(p,a)}_m n \lambda
\over  \cosh {ng\lambda \over 2}},\eqno({\rm B}.12)
$$
where $g$ is the dual Coxeter number.
$z^{(p,a)}_m$ and $\beta^{(p,a)}_m \ge 0$ are the rational constants
uniquely specified by the above expansion.
They satisfy
$$\eqalignno{
z^{(p,a)}_m &> 0,&({\rm B}.13{\rm a})\cr
\pm \beta^{(p,a)}_m - {g \over 2} &\le -1.
&({\rm B}.13{\rm b})\cr}
$$
Here, (B.13b) is a direct
consequence of (B.11b).
The property (B.13a), which is by no means obvious from
the definition, is crucial to assure $z > 0$ etc, in (2.46)
in the main text.
It can be checked
directly by using the explicit forms (B.2-9).
We note that the expansion as (B.12) with the property (B.13a)
is impossible in general without ``averaging over" $Aut$.
The case $X_r = E_6$, $a = p = 1$ is such an example.
\secskip
%
%
%
%

\centerline{\bf Appendix C.  Asymptotic form of
$\hbox{log}YB^{(a)}_m(u)$
for $A_r$}
\vskip0.4cm

Here we shall derive the asymptotic forms (3.12) and (3.37) directly from
the known eigenvalue of the transfer matrix [\rBRT] for
$A_r$ RSOS models [\rJKMO].
We shall calculate it firstly in
the regime $\epsilon = -1$ in the sense of section 3.1.
The behavior in the other regime
$\epsilon = +1$ can then be deduced
from the level-rank duality [\rKNlev].
\par
Let $Tab(\mu)$ denote the set of semi-standard tableaux
on the Young diagram $\mu$ for $sl(r+1)$.
Namely, $t \in Tab(\mu)$ is an assignment of an integer
$t(c,q)$ to each box at position $(c,q)$ in $\mu$ such that
\pn
\item{1.} $1 \le t(c,q) \le r+1$,
\item{2.} $t(c,q) < t(c+1,q)$,
\item{3.} $t(c,q) \le t(c,q+1)$. \par
Then the eigenvalue of transfer matrix with
spectral parameter $u$ is given by [\rBRT]
$$
\Lambda^{\lambda,\mu}(u) = \sum_{t \in Tab(\mu)}
\prod_{(c,q) \in \mu} X^{t(c,q)}(u+\mu_1+c-q-1).\eqno({\rm C}.1)
$$
Here $\lambda$ is the Young diagram
specifying that the transfer matrix
of the corresponding vertex model acts on the quantum space
$V_\lambda^{\otimes N}$.
($V_\lambda$: the irreducible $sl(r+1)$-module with the highest weight
corresponding to $\lambda$.)
Similarly, $\mu$ signifies that the auxiliary space
of the transfer matrix is $V_\mu$.
$\lambda_i\, (\mu_i)$ denotes
the number of boxes in the $i$-th row of $\lambda\, (\mu)$
and
$$\eqalign{
X^c(u)  &=\omega_c {{Q^{c-1}(u-1) Q^{c}(u+1)}
         \over{Q^{c-1}(u) Q^{c}(u)}}f(u-\lambda_c),  \cr
f(u)    &=\sin^N ({\pi \over L} u),  \quad L = \ell + r + 1,  \cr
Q^0(u)  &=Q^{r+1}(u)=1,                                         \cr
Q^c(u) &=\prod_{j=1}^{N_c} \sin({\pi \over L} (u-u^c_j))\quad 1\le c \le r.
\cr }\eqno({\rm C}.2)$$
In the above, the length $N$ of the transfer matrix
is taken as $N \equiv 0$ mod $r+1$ and $N_c$ is given by (IB.1).
$\{\omega_c \}$ are some phase factors [\rBRT] and
$\{u^c_j\}$ are the solutions
to the Bethe ansatz equation (BAE)
$$
{{\omega_c f(u^c_j-\lambda_c)} \over {\omega_{c+1} f(u^c_j-\lambda_{c+1})}}
=
-{{Q^{c+1}(u^c_j+1) Q^c(u^c_j-1) Q^{c-1}(u^c_j)}
\over{Q^{c+1}(u^c_j) Q^c(u^c_j+1) Q^{c-1}(u^c_j-1)}}
\quad \quad 1\le c \le r.
\eqno({\rm C}.3)$$
In the remainder of this appendix, we set
$\lambda =  p \times s,\, \mu = a \times m$ rectangular shapes
in accordance with the setting in section 3.
\par
When $N \rightarrow \infty$, the dominant contribution in (C.1)
is expected to come from the semi-standard tableau such that
$t(c,q) = c, \, \forall q$ as in the case $\lambda = \mu$ [\rBRT].
Being concerned with the bulk behavior, we retain only this term
and thereby get
$$
\log \Lambda^{\lambda,\mu}(u) \simeq
\log {{Q^a(u+m+a-1)}\over {Q^a(u+a-1)}} +
\sum_{c=1}^a \sum_{q=1}^m \log f(u+q+c-2-\lambda_c).
\eqno({\rm C}.4)
$$
Let us rewrite (C.4) in the form suitable for the evaluation at
the ground state in the regime $\epsilon = -1$.
With the change of the variables
$$
u^c_j \rightarrow i w^c_j /2 + (s+c-p)/2,
$$
eq.(C.3) reads ($h(w)=\sinh(\pi w/L)$)
$$\eqalign{{\omega_c \over \omega_{c+1}}
\Biggl( &{{h(w^c_j /2+is\delta_{c p}/2 )}
            \over{h(w^c_j /2-is\delta_{c p}/2)}} \Biggr )^N\cr
= - &\prod_{j'} {{h((w^c_j-w^{c+1}_{j' })/2- i/2 )}
            \over{h((w^c_j-w^{c+1}_{j'}) /2+ i/2 )}}
  \prod_{j''} {{h((w^c_j-w^{c}_{j''})/2+i)}
            \over{h((w^c_j-w^{c}_{j''}) /2-i)}}    \cr
 &\prod_{j'''} {{h((w^c_j-w^{c-1}_{j'''})/2- i/2 )}
            \over{h((w^c_j-w^{c-1}_{j'''}) /2+ i/2 )}},
}\eqno({\rm C}.5)$$
which is nothing but the BAE (IB.2) for $X_r = A_r$.
In the ground state of the regime $\epsilon = -1$ at $N \rightarrow \infty$,
the roots of (C.5) are expected to form the $s$-strings
$w^c_j  =v^c_j + i(s+1-2 \alpha),\,\alpha=1, \cdots, s$.
Then the distribution function of the color$-a$ $s$-string centers
$\{ v^a_j \}$ is given by
$$
{\cal A}^{(r+1)}_{a p}(v),\eqno({\rm C}.6)
$$
defined in (IB.32e) by its Fourier component.
To see this, note that $\forall \sigma^{(a)}_m, \rho^{(a)}_m \equiv 0$
except for $\rho^{(a)}_s$'s in the Dirac sea of
$s$-strings.
Due to (IB.12) and (IB.32e), the BAE (IB.3) then becomes
$\delta_{p a}{\cal A}^{(\ell)}_{s m} =
\sum_{b=1}^r {\cal A}^{(\ell)}_{s m} {\cal M}_{a b}
\rho^{(b)}_s$, which has the solution (C.6)
because of (IB.17a) and (IB.18).
In terms of the $\{ v^a_j \}$, (C.4) is rewritten as
$$\eqalign{
&\log \Lambda^{\lambda,\mu}(u)\cr
&\simeq
\sum_{c=1}^a \sum_{q=1}^m \log f(u+q+c-2-\lambda_c) +
\sum_j \sum_{\alpha=1}^s
\log {{h({v^a_j\over 2}+i(u+m+{a+p-1\over 2})-i\alpha})\over
{h({v^a_j\over 2}+i(u+{a+p-1\over 2})-i\alpha)}}.\cr}
\eqno({\rm C}.7)
$$
Replacing $\sum_j \rightarrow
N\int dv {\cal A}^{(r+1)}_{a p}(v)$ and passing
to the Fourier components,
we get the bulk eigenvalue
$$\eqalign{
\log \Lambda^{\lambda,\mu}_{{\rm bulk}}(u) =
&\sum_{c=1}^a \sum_{q=1}^m \log f(u+q+c-2-\lambda_c) \cr
&+ N \int dk {{\sinh (2u+m-s+a+p-2)k}\over{k}}
{\hat {\cal A}}^{(L)}_{s m}(k) {\hat {\cal A}}^{(r+1)}_{a p}(k).}
\eqno({\rm C}.8)
$$
The eigenvalue of the $T^{(a)}_m(u)$ in section 3
is given by
$\Lambda^{\lambda, \mu}({u \over 2i} + {s-p+{r+1 \over 2}-m-a\over 2})$
here.
Thus the bulk term for the $Y$-system (3.31) is given by
$$
YB^{(a)}_m(u) =
{{\Lambda^{\lambda,\mu^{(1)}}_{{\rm bulk}}
({u \over 2i} + {s-p+{r+1 \over 2}-m-a-1\over 2})
\Lambda^{\lambda,\mu^{(2)}}_{{\rm bulk}}
({u \over 2i} + {s-p+{r+1 \over 2}-m-a+1\over 2})  }
             \over
{{\Lambda^{\lambda,\mu^{(3)}}_{{\rm bulk}}
({u \over 2i} + {s-p+{r+1 \over 2}-m-a-1\over 2})
\Lambda^{\lambda,\mu^{(4)}}_{{\rm bulk}}
({u \over 2i} + {s-p+{r+1 \over 2}-m-a+1\over 2})  } }
},\eqno({\rm C}.9)
$$
where $\mu^{(1)}, \ldots, \mu^{(4)}$ are the Young diagrams with the
rectangular shapes
$a\times (m+1), a\times (m-1),(a+1)\times m,(a-1)\times m$,
respectively.
Here we have used the facts $g^{(a)}_m(u)=1 \,(1 \le a \le r-1)$ and
$g^{(r)}_m(u)=
\Lambda^{\lambda, \mu'}({u \over 2i} + {s-p-{r+1 \over 2}-m\over 2})$
($\mu' = r \times m$ Young diagram) in (3.31).
These are derivable from section 2.3 in Part I and
eq.(3.12) in [\rBRT].
Substitute (C.8) into the logarithm of (C.9) and
simplify the result by means
of the formulas that follow from the definition (IB.32e)
$$\eqalignno{
&\hat{\cal A}^{(T)}_{a\, b+1}(k) + \hat{\cal A}^{(T)}_{a\, b-1}(k)
= 2\cosh k \hat{\cal A}^{(T)}_{a b}(k) - \delta_{a b},
&({\rm C}.10{\rm a})\cr
&\int dk{\sinh(2u+b-a+1)k \over k}\hat{\cal A}^{(T)}_{a b}(k) =
- \sum_{j=1}^b \log {\sin \bigl({\pi\over T}(u-a+j)\bigr) \over
\sin \bigl({\pi\over T}(u+j)\bigr)},
&({\rm C}.10{\rm b})\cr
}$$
which is valid for any $a,b \in \{1,2,\ldots,T-1\},
T \in {\bf Z}_{\ge 2}$.
After some calculations we obtain
$$
\log YB^{(a)}_m(u) = N \delta_{s\,m} \sum_{c=0}^{p-1}
\log {
\sin\bigl({\pi\over r+1}
({u \over 2i} + {-p+{r+1 \over 2}-a-1\over 2} + c)\bigr)
\over
\sin\bigl({\pi\over r+1}
({u \over 2i} + {-p+{r+1 \over 2}+a-1\over 2} + c)\bigr)
}, \eqno({\rm C}.11)
$$
which indeed satisfies (3.35).
{}From this one can easily derive the
$N \rightarrow \infty$ asymptotic form
in the regime $\epsilon = -1$
$$\eqalign{
&\log YB^{(a)}_m(\pm(u+{r+1\over \pi}\log N)) \cr
&= \delta_{s m}\Bigl(\mp{apN\pi i\over r+1}
-2{\sin{\pi a\over r+1}\sin{\pi p\over r+1}\over\sin{\pi \over r+1}}
e^{-{\pi u\over r+1}} + O({1\over N})\Bigr).\cr}
\eqno({\rm C}.12)
$$
To get the behavior in the other regime $\epsilon = +1$,
we invoke the level-rank duality [\rKNlev].
Noting that the rhs of (C.12) is symmetric under
$a \leftrightarrow p$ and $m \leftrightarrow s$, we
formally exchange $r+1 \leftrightarrow \ell$,
$a \leftrightarrow m$ and $p \leftrightarrow s$ there
to get
$$\eqalign{
&-\log YB^{(a)}_m(\pm(u+{\ell\over \pi}\log N)) \cr
&= \delta_{p a}\Bigl(\mp{msN\pi i\over \ell}
-2{\sin{\pi m\over \ell}\sin{\pi s\over \ell}\over\sin{\pi \over \ell}}
e^{-{\pi u\over \ell}} + O({1\over N})\Bigr)\cr}
\eqno({\rm C}.13)
$$
in the regime $\epsilon = +1$.
Here the overall sign factor on the lhs has been attached to take
the difference of (3.6) and (3.34) into account.
When $r=1$, (C.13) indeed agrees, with a suitable adjustment
of the variables, with the regime II
result eq.(4.32) in [\rKPT].
\secskip
%
%
%

\centerline{\bf Appendix D. Rogers dilogarithm and the conjecture in [\rKN]}
\vskip0.6cm

This appendix is devoted to a description of the Rogers dilogarithm
and the curious conjecture
proposed in [\rKN].
The latter relates the parafermion scaling dimensions
to a certain sum of the dilogarithm special values.
As promised there we present here some interesting calculations
that elucidate the origin of the conjecture.
In section D.1 we recall the log and the dilogarithm
functions and specify our convention on their analytic continuations.
Our dilogarithm conjecture will be formulated in D.2
following [\rKN] for any classical simple Lie algebra
$X_r$, integer $\ell \ge 1$ and dominant integral weight
$\Lambda \in P_\ell$.
(See (I3.4) for the definition of $P_\ell$.)
Its special case $\Lambda = 0$ concerns the earlier
one in [\rKi,\rKu] and is relevant to our calculation in
section 3 of the main text.
(Conversely we expect that the content of section 3
could be extended so that the $\Lambda$-generic ($\in P_\ell$)
dilogarithm conjecture becomes relevant.)
Section D.3 contains
a proof of the important property eq.(16) in [\rKN] based on
the congruence index explored in appendix A of Part I.
We remark that the connection of the dilogarithm
with CFT has attracted much attention
recently and several insights have been gained in
[\rKNS,\rKNC-\rDuSa].
\subskip
\par\noindent
{\bf D.1. $\log f$, $L(f)$ and their analytic continuations}
\par
Let $\log x$ signify the logarithm in the branch
$-\pi < {\rm Im}(\log x) \le \pi$ for $x \neq 0$.
Namely,
$\log x = \log \vert x \vert + i \pi\arg(x)$ with
$-1 < \arg(x) \le 1$ for any non-zero
$x = \vert x \vert e^{i\pi\arg(x)}$.
Under this convention for the $\arg$ function, we have
$$
\log(x_1^{r_1}x_2^{r_2} \cdots x_k^{r_k}) =
r_1 \log x_1 + \cdots + r_k \log x_k - 2\pi i
H(r_1\arg(x_1)+ \cdots + r_k\arg(x_k)), \eqno({\rm D}.1{\rm a})
$$
for any nonzero complex numbers $x_1, \ldots, x_k$ and
real numbers $r_1, \ldots, r_k$.
The function
$H(x) \in {\bf Z}$ for $x \in {\bf R}$ is given by the rule
$$H(x) = n\,\, \hbox{ for }\,\, 2n-1 < x \le 2n+1, \,\,
n \in {\bf Z}.
\eqno({\rm D}.1{\rm b})
$$
By the definition, the functions
$\log x$ and $\log (1-x)$ have the branch cuts $(-\infty,0]$ and
$[1,\infty)$ on the real $x$-axis and they actually belong to
the upper and lower half plane, respectively.
The Rogers dilogarithm is a function
of a complex number $f$ defined by
$$
L(f) = -{1 \over 2} \int_0^f
\Bigl( {\log (1-x) \over x} + { \log \, x \over 1 - x }\, \Bigr) dx,
\eqno({\rm D}.2)
$$
where the integral is along a contour that does not cross the
cuts mentioned above.
\par
To discuss the analytic continuations of these functions,
we introduce the universal covering space
${\cal R}$ of ${\bf C}\setminus \{0,1\}$ and the covering map
${\tilde i}: {\cal R} \rightarrow {\bf C}\setminus \{0,1\}$.
Given $f \in {\bf C}\setminus \{0,1\}$,
a corresponding point ${\tilde f} \in {\cal R}$,
${\tilde i}({\tilde f}) = f$
is specified
by a contour ${\cal C}$ from an arbitrarily fixed base point to $f$
in ${\bf C}\setminus \{0,1\}$.
Since ${\tilde f}$ depends only on the homotopy class of
${\cal C}$, we shall parameterize it by the integers
$\xi_j, \eta_j\, (j \ge 1)$ as
${\cal C} = {\cal C}[f \vert
\xi_1,\xi_2, \ldots \vert
\eta_1,\eta_2, \ldots]$, which signifies the following contour.
It firstly
goes across the cut $[1,\infty)$ for $\eta_1$ times then
crosses the other cut $(-\infty, 0]$
for $\xi_1$ times then
$[1,\infty)$ again for $\eta_2$ times,
$(-\infty, 0]$ for $\xi_2$ times and so on
before approaching $f$ finally.
Here intersections have been counted as $+1$ when
the contour goes across the cut $(-\infty, 0]$ (resp. $[1,\infty)$)
from the upper (resp. lower) half plane to the lower
(resp. upper) and $-1$ if opposite.
We call $\xi_j$ and $\eta_j$ the winding numbers and
assume that they are all
zero for $j$ sufficiently large.
Let ${\tilde f} \in {\cal R}$ be so specified
by ${\cal C}$ and let ${\rm Log}{\tilde f}$ and
${\tilde L}({\tilde f})$ denote the analytic continuations
of $\log f$ and $L(f)$ to ${\cal R}$, respectively.
{}From the definitions one deduces the
formulas
$$\eqalignno{
\Log(\tilde{f}) &= \log f + 2\pi i (\sum_{j \ge 1}\xi_j),\quad
\Log(1-\tilde{f}) = \log (1-f) + 2\pi i (\sum_{j \ge 1}\eta_j),
&({\rm D}.3{\rm a})\cr
\tilde{L}(\tilde{f}) &= L(f) +
\pi i \bigl(\sum_{j \ge 1}\xi_j \bigr) \log (1-f) -
\pi i \bigl(\sum_{j \ge 1}\eta_j \bigr) \log f \cr
& \quad
- 2\pi^2\bigl(\sum_{j \ge 1}\xi_j \bigr)
\bigl(\sum_{j \ge 1}\eta_j \bigr) + 4\pi^2 \sum_{j \ge 1}
\xi_j(\eta_1 + \cdots + \eta_j), &({\rm D}.3{\rm b})\cr}
$$
which make the dependences on the contour
${\cal C} = {\cal C}[f \vert
\xi_1,\xi_2, \ldots \vert
\eta_1,\eta_2, \ldots]$ explicit.
We note that eq.(9b) and (12d) in [\rKN] are erroneous
(though it does not affect the conclusion there) and should be corrected
as (D.3b) and (D.12d) in this paper, respectively.
\par
\vskip0.3cm\noindent
{\bf D.2. Dilogarithm conjectures}

\par
In the remainder of this appendix, we fix an integer
$\ell \ge 1$ and use the notation (3.1b).
Define a complex valued function $f^{(a)}_m(z)$ on
the dual space of the Cartan subalgebra $z \in {\cal H}^\ast$ by
$$
f^{(a)}_m(z) = 1 -
{Q^{(a)}_{m-1}(z)Q^{(a)}_{m+1}(z) \over Q^{(a)}_m(z)^2}\quad
{\rm for }\,\, (a,m) \in G,
\eqno({\rm D}.4)
$$
where the quantity $Q^{(a)}_m(z)$ has been detailed in
section 3.2 and appendix A of Part I.
Among other values of $z$,
$Q^{(a)}_m(0)$ is real positive for all
$(a,m) \in G$ as noted after (I3.8).
As shown in (IB.28), it follows from (D.4) that
$$\eqalignno{
&0 < f^{(a)}_m(0) < 1\quad \hbox{for }\,\, (a,m) \in G,
&({\rm D}.5{\rm a})\cr
&\log f^{(a)}_m(0) = \sum_{(b,k) \in G} K_{a b}^{m k}
\log \bigl(1-f^{(b)}_k(0)\bigr),
&({\rm D}.5{\rm b})\cr}
$$
where $K_{a b}^{m k}$ is defined in (IB.21) and
related to $J_{b a}^{k m}$ in (I3.7), (IB.22) via (IB.23a).
It has been conjectured in [\rKi,\rKu] that
$$
{6 \over \pi^2}\sum_{(a,m) \in G} L\bigl(f^{(a)}_m(0)\bigr)
= {\ell \hbox{ dim } X_r \over \ell + g} - r.\eqno({\rm D}.6)
$$
The rhs is the well known
level $\ell$ $X^{(1)}_r$ parafermion central charge
$c_{\rm PF}$ [\rGe]
while the intricate lhs arises both from
the $T$-system analysis in section 3 and
the thermodynamic Bethe ansatz [\rKu-\rBRT].
So far the proof of (D.6) has been known
for $X_r = A_r$ in [\rKi,\rKRjp] as well as several insights
[\rKNS,\rKNC-\rDuSa] for related problems.
In particular, a new $q$-series formula has been conjectured
in [\rKNS] for a level $\ell$ $X^{(1)}_r$ string function [\rKPet],
which may be viewed as a $q$-analogue of (D.6).
\par
Let us proceed to the generalization [\rKN] of (D.6) where the rhs
includes the combination
$c_{\rm PF} - 24 \times$(scaling dimensions mod ${\bf Z}$).
Following [\rKN], we call an element
$z \in {\cal H}^\ast$ {\it regular} if it satisfies
$Q^{(a)}_m(z) \neq 0$ in (I3.6) for all
$1 \le a \le r, 1 \le m \le \ell_a$
and {\it singular} otherwise.
The most typical example of the former is
the choice $z=0$ since all the relevant $Q^{(a)}_m(0)$'s are real positive.
Suppose $z \in {\cal H}^\ast$ is regular.
Then $f^{(a)}_m(z)$ (D.4) is finite and
$f^{(a)}_m(z) \neq 0, 1$ for all $(a,m) \in G$.
Take a contour
${\cal C}_{a,m}
= {\cal C}[f^{(a)}_m(z) \vert \xi^{(a)}_{m,1}, \xi^{(a)}_{m,2},\ldots
\vert \eta^{(a)}_{m,1}, \eta^{(a)}_{m,2},\ldots]$
and denote by ${\tilde f}^{(a)}_m(z)$ the point on
${\cal R}$ specified by the ${\cal C}_{a,m}$ with
${\tilde i}({\tilde f}^{(a)}_m(z)) = f^{(a)}_m(z)$.
Then the collection
$({\tilde f}^{(a)}_m(z))_{(a,m) \in G}$
may be viewed as a point
on ${\cal R}^{\vert G \vert}$ parameterized by
a regular $z \in {\cal H}^\ast$ and
the winding numbers
${\cal S} = (\xi^{(a)}_{m,j},\eta^{(a)}_{m,j})_{(a,m)\in G,\, j \ge 1}$.
According to eq.(11) in [\rKN] we introduce the
following functions of regular $z$ and ${\cal S}$,
$$\eqalignno{
&{\pi^2 \over 6}c(z,{\cal S}) =
\sum_{(a,m) \in G}\Bigl(
{\tilde L}\bigl({\tilde f}^{(a)}_m(z)\bigr) -
{\pi i\over 2}D^{(a)}_m(z,{\cal S})
\Log\bigl(1-{\tilde f}^{(a)}_m(z)\bigr)\Bigr),&({\rm D}.7{\rm a})\cr
&\pi i D^{(a)}_m(z,{\cal S}) =
\Log\bigl({\tilde f}^{(a)}_m(z)\bigr) -
\sum_{(b,k) \in G}K^{m\, k}_{a\, b}
\Log\bigl(1-{\tilde f}^{(b)}_k(z)\bigr).&({\rm D}.7{\rm b})\cr
}$$
Note that for the typical regular element $z = 0 \in {\cal H}^\ast$,
(D.7) provides an analytic continuation of the
$u = +\infty$ term in (3.25) due to (3.26).
In particular, if all the winding numbers are zero further,
$\hbox{Log} {\tilde f}^{(a)}_m(0)$ becomes
$\log f^{(a)}_m(0)$ itself and (D.7b) vanishes
because of (D.5b).
Therefore the simplest case $c(z=0,{\cal S}=\hbox{trivial})$
of (D.7a) is nothing
but the lhs of (D.6).
Thus we expect, in the light of the calculation in section 3.2.4,
that the function $c(z,{\cal S})$ will be relevant to
the combination
$c_{{\rm eff}} = c_{\rm PF} - 24\times(\hbox{ scaling dimension})$ and
possibly leads to a generalization of (D.6).
This seems indeed the case as will be formulated in our conjecture shortly.
We remark that there is yet another route to reach (D.7) from the
TBA argument as done in [\rKNS].
These are the two sources mentioned in [\rKN]
that led to the invention of (D.7).
\par
Given a regular $z \in {\cal H}^\ast$, define an element
$\lambda(z) \in {\cal H}^\ast$ by (cf.\ eq.(13) in [\rKN])
$$\eqalignno{
\lambda(z) &= \sum_{a=1}^r \lambda_a(z) \Lambda_a,
&({\rm D}.8{\rm a})\cr
\lambda_a(z) &= {1 \over 2 \pi i}\sum_{b=1}^r C_{a b}
\Bigl(\sum_{k=1}^{\ell_b-1} k \log(1-f^{(b)}_k(z)) &\cr
&+ \ell_b\bigl(
\log Q^{(b)}_{\ell_b-1}(z) -
\log Q^{(b)}_{\ell_b}(z)\bigr)\Bigr),&({\rm D}.8{\rm b})\cr}
$$
where $\log$ denotes the logarithm specified in section D.1.
{}From now on we shall mainly concern the
specialization to the dominant integral weights
$z = \Lambda \in P_\ell$ (I3.4).
In case $\Lambda$ is singular, formulas are to be understood
via a fixed way of the limit
$z \in {\cal H}_{\bf R}^\ast \rightarrow \Lambda$.
Based on (IB.23) and (IA.11), it will be shown in Proposition 1 of
section D.3 that
$$\lambda(\Lambda) \equiv \Lambda\quad \hbox{ mod }\,Q,
\eqno({\rm D}.9)
$$
where $Q$ stands for the root lattice as in section 3.1 of Part I.
This is eq.(16a) of [\rKN].
Given the winding numbers
${\cal S} = (\xi^{(a)}_{m,j}, \eta^{(a)}_{m,j})_{
(a,m) \in G, j \ge 1}$,
we put
$$\eqalignno{
\beta({\cal S}) &= \sum_{(a,m) \in G} m N^{(a)}_m \alpha_a \in Q,
&({\rm D}.10{\rm a})\cr
N^{(a)}_m &= \sum_{j \ge 1} \eta^{(a)}_{m,j} \in {\bf Z},
&({\rm D}.10{\rm b})\cr}
$$
in which (D.10b) is finite since only finitely many
$\eta^{(a)}_{m,j}$'s are non-zero as noted
before (D.3).
Now the generalized dilogarithm conjecture
can be stated as follows.
\proclaim Conjecture (eq.(17) of [\rKN]).
Let $c(z, {\cal S})$ be as defined in (D.7) and take
a dominant integral weight $\Lambda \in P_\ell$.
Then,
$$\eqalignno{
&c(\Lambda,{\cal S}) = {\ell {\rm dim }X_r \over \ell + g} - r
- 24(\Delta^\Lambda_{\lambda(\Lambda)+\beta({\cal S})} +
{\rm integer}),&({\rm D}.11{\rm a})\cr
&\Delta^z_y = {(z \vert z+2\rho) \over 2(\ell + g)} -
{\vert y \vert^2 \over 2\ell}\quad{\rm for }\,\,
y, z \in {\cal H}^\ast,&({\rm D}.11{\rm b})\cr}
$$
where $\rho$ stands for the half sum of the positive roots of
$X_r$ (I3.1e).

The rhs of (D.11a) with the congruence
properties (D.9) and (D.10a) is well known as the
central charge $- 24\times$ (scaling dimension mod ${\bf Z}$) of the
level $\ell \,\, X^{(1)}_r$ parafermion CFT [\rGe].
As the winding number ${\cal S}$ is chosen variously,
$\beta({\cal S})$ (D.10a)
ranges over the root lattice $Q$.
Thus all the spectra in parafermion CFTs seem to come out
from the Rogers dilogarithm through the quantity
$c(\Lambda,{\cal S})$ (D.7), which was the main observation in [\rKN].
In fact, when $\Lambda = 0 \in P_\ell$,
a more concrete correspondence
can be inferred between vacuum parafermion module and
the contours represented by ${\cal S}$ as argued in [\rKNS].
\par
In the rest of this section, we rewrite our conjecture
in a simpler form suitable for the discussion in section D.3.
Firstly, we extract the ${\cal S}$-dependence of
the quantity $c(\Lambda,{\cal S})$ by applying the formula
(D.3).
A little manipulation leads to
$$\eqalignno{
c(\Lambda,{\cal S})
&= c_0(\Lambda) - 24T(\Lambda,{\cal S}),&({\rm D}.12{\rm a})\cr
{\pi^2 \over 6}c_0(\Lambda) &=
\sum_{(a,m) \in G}\Bigl(
L\bigl(f^{(a)}_m(\Lambda)\bigr) -
{\pi i\over 2}d^{(a)}_m(\Lambda)
\log\bigl(1-f^{(a)}_m(\Lambda)\bigr)\Bigr),&({\rm D}.12{\rm b})\cr
\pi i d^{(a)}_m(\Lambda) &=
\log f^{(a)}_m(\Lambda) -
\sum_{(b,k) \in G}K^{m\, k}_{a\, b}
\log\bigl(1-f^{(b)}_k(\Lambda)\bigr),&({\rm D}.12{\rm c})\cr
T(\Lambda,{\cal S}) &= {1 \over 2}
\sum_{\scriptstyle (a,m) \in G \atop \scriptstyle (b,k) \in G}
K^{m\, k}_{a\, b} N^{(a)}_m N^{(b)}_k
- {1 \over 2}\sum_{(a,m) \in G}d^{(a)}_m(\Lambda) N^{(a)}_m \cr
&\qquad - \sum_{(a,m) \in G}\sum_{j \ge 1}
\xi^{(a)}_{m,j}(\eta^{(a)}_{m,1} + \cdots + \eta^{(a)}_{m,j}),
&({\rm D}.12{\rm d})\cr}$$
where $N^{(a)}_m$ is specified from
${\cal S} = (\xi^{(a)}_{m,j},\eta^{(a)}_{m,j})_{(a,m)\in G,\, j \ge 1}$
by (D.10b).
Here comes the point which enabled us to find the conjecture
(D.11).
If one expects the parafermion scaling dimension from
$c(\Lambda,{\cal S})$ at all, there must be a square of
some vector corresponding to $-{\vert y \vert^2 \over 2\ell}$ in
(D.11b) such that $y \equiv \Lambda$ mod $Q$.
Our idea is to regard $T(\Lambda,{\cal S})$ (D.12d)
as a quadratic form of $N^{(a)}_m \in {\bf Z}$ and
try to complete a square out of it mod ${\bf Z}$.
Actually this postulate was almost enough to find the vector
$\lambda(\Lambda)$ (D.8) for which eq.(16b) in [\rKN], i.e.,
$$
T(\Lambda,{\cal S}) \equiv -{1 \over 2\ell}
\vert \lambda(\Lambda) + \beta({\cal S}) \vert^2 +
{1 \over 2\ell}\vert \lambda(\Lambda) \vert^2 \quad
{\rm mod }\,\, {\bf Z}
\eqno({\rm D}.13)
$$
certainly holds.
See Proposition 2 in section D.3 for the proof.
In view of (D.13) and (D.12a) the conjecture (D.11) reduces to a
simpler statement (eq.(14) in [\rKN])
$$
c_0(\Lambda) = {\ell{\rm dim}X_r \over \ell + g} - r -
24(\Delta^\Lambda_{\lambda(\Lambda)} + {\rm integer}),
\eqno({\rm D}.14)
$$
for $c_0(\Lambda)$ defined by (D.12b,c).
We have checked this numerically for all the regular
$\Lambda \in P_\ell$ in
$(X_r,\ell) = (A_1, 2\sim15), (A_2,2\sim6), (A_3,2\sim4),
(A_{4,5},2\sim3),(B_r,6-r),(C_r,6-r)$ for $2 \le r \le 5,
(D_{4,5},2\sim3)$ and $(D_6,2)$.
Singular $\Lambda$ case has also been confirmed in the above
examples under several numerical limits
$z (\in  {\cal H}^\ast_{\bf R}) \rightarrow \Lambda$.
We note that when $\Lambda = 0$,
$d^{(a)}_m(0) = \lambda(0) = 0$ holds, therefore the above
conjecture (with ``integer" = 0) reduces to (D.6).
\par\vskip0.3cm
\noindent
{\bf D.3. Proof of (D.9) and (D.13)}

\par
Let us show the properties (D.9) and (D.13) based on
(IA.11).
They have been announced earlier in
eqs.(16a) and (16b) of [\rKN], respectively.
We shall exclusively
consider the specialization $z = \Lambda \in P_\ell$.
As mentioned after (D.8),
formulas involving $\Lambda$ should then be understood via a limit
$z \in {\cal H}_{\bf R}^\ast \rightarrow \Lambda$
whenever $\Lambda$ is singular.
For simplicity we often abbreviate $Q^{(a)}_m(\Lambda)$
to $Q^{(a)}_m$ etc and use the notation
$\theta^{(a)}_m = \hbox{ arg }Q^{(a)}_m(\Lambda)$.
Thus we have
$$
Q^{(a)}_m(\Lambda) = \vert Q^{(a)}_m(\Lambda) \vert
\hbox{exp}(i\pi \theta^{(a)}_m), \quad
-1 < \theta^{(a)}_m \le 1,\eqno({\rm D}.15)
$$
for $ 1 \le a \le r, \, 0 \le m \le \ell_a$ in accordance with
section D.1.
\par
Our first step is to rewrite the $d^{(a)}_m$ (D.12c) as
\proclaim Lemma D.1. For $(a,m) \in G$ we have
$$\eqalignno{
&d^{(a)}_m = -{m \over \ell_a \pi i}
\sum_{b=1}^r C_{a b} \log Q^{(b)}_{\ell_b}
+ 2\sum_{(b,k) \in G}K^{m k}_{a b}
H(\theta^{(b)}_{k+1}+\theta^{(b)}_{k-1}-2\theta^{(b)}_k)\cr
&\qquad-2H(-2\sum_{b=1}^r\sum_{k=1}^{\ell_b}J^{k m}_{b a}\theta^{(b)}_k),
&({\rm D}.16{\rm a})\cr
&d^{(a)}_m \equiv md^{(a)}_1 \quad \hbox{ mod } 2{\bf Z}.
&({\rm D}.16{\rm b})\cr}
$$

\noindent
{\it Proof.}
Noting $Q^{(b)}_0 = 1$ but $Q^{(b)}_{\ell_b} \neq 1$ in general,
the definition (D.4) can be written as
$$1 - f^{(b)}_k = Q^{(b)\, \delta_{k \ell_b - 1}}_{\ell_b}
\prod_{n=1}^{\ell_b - 1}
Q_n^{(b) \, -{\bar C}^b_{n k}}, \eqno({\rm D}.17{\rm a})
$$
for $(b,k) \in G$, where ${\bar C}^b$ denotes the Cartan matrix of
$A_{\ell_b - 1}$ as in (IB.16b).
Combining (D.4) with (IA.9a) we get
$$
f^{(a)}_m = \prod_{(b,k) \in G}
Q^{(b) \, -2J^{k m}_{b a}}_k
\prod_{b=1}^r
Q^{(b) \, -2J^{\ell_b  m}_{b a}}_{\ell_b}, \eqno({\rm D}.17{\rm b})
$$
for $(a,m) \in G$, where we have extracted the factor corresponding to
$k = \ell_b$ in (IA.9a) explicitly.
Expanding the $\log$ of (D.17a)
according to the rule (D.1a) gives
$$
\log(1-f^{(b)}_k) = \delta_{k \, \ell_b - 1} \log Q^{(b)}_{\ell_b}
-\sum_{n=1}^{\ell_b-1}{\bar C}^b_{n k}\log Q^{(b)}_n
-2\pi i H(\theta^{(b)}_{k+1}+\theta^{(b)}_{k-1}-2\theta^{(b)}_k).
\eqno({\rm D}.18)
$$
Substitute this and the similar expression for
$\log f^{(a)}_m$ obtainable from (D.17b) into (D.12c).
The result reads
$$\eqalign{
&\pi i d^{(a)}_m\cr
&= -2 \sum_{(b,k) \in G} J^{k m}_{b a}
\log Q^{(b)}_k - 2\sum_{b=1}^rJ^{\ell_b m}_{b a}
\log Q^{(b)}_{\ell_b}
-2\pi i H(-2\sum_{b=1}^r
\sum_{k=1}^{\ell_b}J^{k m}_{b a}\theta^{(b)}_k) \cr
-&\sum_{(b,k) \in G}K^{m k}_{a b}\Bigl[
\delta_{k \, \ell_b - 1} \log Q^{(b)}_{\ell_b}
-\sum_{n=1}^{\ell_b-1}{\bar C}^b_{n k}\log Q^{(b)}_n
-2\pi i H(\theta^{(b)}_{k+1}+\theta^{(b)}_{k-1}-2\theta^{(b)}_k)
\Bigr],\cr
}\eqno({\rm D}.19)$$
which consists of 6 pieces.
The first one cancels the fifth due to (IB.23a).
The second one can be combined with the fourth by
means of (IB.23c), which proves (D.16a).
To show (D.16b), recall from the definition (D.1b)
that $H(\cdot)$'s in (D.16a) are just integers.
Using the explicit forms (IB.21b) and
$(\alpha_a \vert \alpha_b) = C_{a b}/t_a$ further, one finds
$$
d^{(a)}_m \equiv -{m \over \ell_a\pi i}
\sum_{b=1}^rC_{a b}\log Q^{(b)}_{\ell_b} -
{2m \over \ell_a}\sum_{(b,k) \in G}C_{a b}k
H(\theta^{(b)}_{k+1}+\theta^{(b)}_{k-1}-2\theta^{(b)}_k)\,
\hbox{ mod } 2{\bf Z},
\eqno({\rm D}.20)
$$
from which (D.16b) follows.
\par
Similarly $\lambda_a(\Lambda)$ (D.8b) is expressed as
\proclaim Lemma D.2. For $1 \le a \le r$,
$$
\lambda_a(\Lambda) =
-{1 \over 2 \pi i}
\sum_{b=1}^r C_{a b} \log Q^{(b)}_{\ell_b}(\Lambda)
- \sum_{(b,k) \in G} C_{a b}k
H(\theta^{(b)}_{k+1}+\theta^{(b)}_{k-1}-2\theta^{(b)}_k),
\eqno({\rm D}.21{\rm a})$$
$$\eqalignno{
&d^{(a)}_1(\Lambda) \equiv {2 \over \ell_a}\lambda_a \quad
\hbox{ mod } 2{\bf Z},&({\rm D}.21{\rm b})\cr
&\lambda_a(\Lambda) \in {\bf Z}.&({\rm D}.21{\rm c})\cr
}$$

\noindent
{\it Proof.}
To see (D.21a), replace $\log(1-f^{(b)}_k)$ in (D.8b) with the rhs of
(D.18) and reduce the result by
$\sum_{k=1}^{\ell_b - 1}k {\bar C}^b_{n k} = \ell_b
\delta_{n \, \ell_b - 1}$.
Comparing (D.21a) and (D.20) we get (D.21b).
Eq.(D.21c) is valid if the first term on the rhs of (D.21a)
is an integer.
{}From the data (IA.11), it can be computed explicitly as
(when $X_r \neq D_r$ for example)
$$
-{1 \over 2\pi i} \sum_{b=1}^rC_{a b}
\Bigl(-{2\pi i \Gamma(\Lambda){\overline \gamma}_b \over \kappa}
- 2\pi i H(-{2\Gamma(\Lambda){\overline \gamma}_b\over\kappa})\Bigr)
\equiv {\Gamma(\Lambda) \over \kappa}
\sum_{b=1}^r C_{a b}{\overline \gamma}_b \equiv 0 \,
\hbox{ mod } {\bf Z},\eqno({\rm D}.22)
$$
where we have used (D.1a), (IA.14) and the fact
$\Gamma(\Lambda), H(\cdot) \in {\bf Z}$.
The case $X_r = D_r$ is verified similarly.
This completes the proof of (D.21c), hence Lemma D.2.
\par
We remark that
$$
d^{(a)}_m(\Lambda) \equiv {2m\lambda_a(\Lambda) \over \ell_a}
\quad \hbox{ mod } \, 2{\bf Z},\eqno({\rm D}.23)
$$
due to (D.16b) and (D.21b).
{}From (D.21c) we now know that $\lambda(\Lambda)$
is an integral weight.
Moreover it has a remarkable property as follows.
\proclaim Proposition 1 (eq.(D.9)).
$\lambda(\Lambda)$ and $\Lambda$ are congruent with respect to
the root lattice $Q$.

\noindent
{\it Proof.}
We consider the case $X_r \neq D_r$.
The case $X_r = D_r$ is similar.
In view of (IA.13) it is sufficient to check
$\Gamma(\lambda(\Lambda)) \equiv \Gamma(\Lambda)$ mod $\kappa{\bf Z}$.
Substitute (D.21a) into the sum
$\Gamma(\lambda(\Lambda)) = \sum_{a=1}^r \gamma_a\lambda_a(\Lambda)$
(IA.12a).
Because of (IA.14) the second term on the rhs of (D.21a)
makes no contribution mod $\kappa{\bf Z}$.
Thus from (D.21a) and (IA.11) we have
$$\eqalign{
\Gamma(\lambda(\Lambda)) &\equiv
-{1 \over 2 \pi i}\sum_{1 \le a, b \le r}
\gamma_a C_{a b}\log Q^{(b)}_{\ell_b}(\Lambda) \,\,
\hbox{ mod } \kappa{\bf Z} \cr
&\equiv
-{1 \over 2\pi i} \sum_{1 \le a, b \le r} \gamma_a C_{a b}
\Bigl(-{2\pi i \Gamma(\Lambda){\overline \gamma}_b \over \kappa}
- 2\pi i H(-{2\Gamma(\Lambda){\overline \gamma}_b\over\kappa})
\Bigr)  \,\, \hbox{ mod } \kappa{\bf Z}\cr
&\equiv \Gamma(\Lambda) \,\, \hbox{ mod } \kappa{\bf Z}, \cr}
\eqno({\rm D}.24)$$
where in the last step we used (IA.14) and (IA.15a).
\par
Eq.(D.9) is now proved.
The significance of $\lambda(\Lambda)$
is not only the congruence property
$\lambda(\Lambda) \equiv \Lambda \,\hbox{ mod } Q$
established above.
It emerges when one tries to complete a square out of the
$T(\Lambda, {\cal S})$ (D.12d) as a quadratic form of
$\{ N^{(a)}_m \mid (a,m) \in G\}$ mod ${\bf Z}$.
Quoting (D.13) and (D.10a) again we have
\proclaim Proposition 2 (eq.(D.13)).
$T(\Lambda, {\cal S})$ defined in (D.12d) satisfies
$$\eqalignno{
T(\Lambda, {\cal S}) &\equiv
-{1 \over 2\ell}
\vert \lambda(\Lambda) + \beta({\cal S})\vert^2
+ {1 \over 2\ell}
\vert \lambda(\Lambda) \vert^2 \, \,\hbox{ mod } {\bf Z},
&({\rm D}.25{\rm a})\cr
\beta({\cal S}) &= \sum_{a=1}^r\sum_{m=1}^{\ell_b - 1}
mN^{(a)}_m \alpha_a \in Q.&({\rm D}.25{\rm b})\cr
}$$

\noindent
{\it Proof.}
Consider the first term
$${1 \over 2}\sum K^{m k}_{a b}N^{(a)}_m N^{(b)}_k =
{1 \over 2}\sum\bigl(\hbox{min}(t_b m, t_a k)
- {m k\over \ell}\bigr)(\alpha_a \vert \alpha_b)N^{(a)}_m N^{(b)}_k$$
appearing in $T(\Lambda, {\cal S})$ (D.12d) (see (IB.21b)).
The $\hbox{min}(\,,\,)$ part
here can be ignored mod ${\bf Z}$.
As for the
$-{1 \over 2}\sum d^{(a)}_m(\Lambda)N^{(a)}_m$ term in (D.12d),
one can replace the $d^{(a)}_m(\Lambda)$ by
${2m \over \ell_a}\lambda_a(\Lambda)$ mod ${\bf Z}$
thanks to (D.23).
The result now reads
$$\eqalign{
T(\Lambda,{\cal S}) \equiv
&-{1 \over 2\ell}\sum_{1 \le a,b \le r}
(\alpha_a \vert \alpha_b)
\Bigl(\sum_{m=1}^{\ell_a - 1}mN^{(a)}_m \Bigr)
\Bigl(\sum_{k=1}^{\ell_b - 1}kN^{(b)}_k \Bigr)
\cr
&-{1 \over \ell}\sum_{a=1}^r
{\lambda_a(\Lambda) \over t_a}
\Bigl(\sum_{m=1}^{\ell_a - 1}mN^{(a)}_m \Bigr)\,\,
\hbox{ mod } {\bf Z}.\cr
}\eqno({\rm D}.26)$$
Noting (D.25b) and
$(\lambda(\Lambda) \vert \alpha_a) = \lambda_a(\Lambda)/t_a$,
we find that the rhs of (D.26) equals
$-{1 \over 2\ell}\vert \beta({\cal S}) \vert^2 -
{1 \over \ell}\bigl(\lambda(\Lambda) \vert \beta({\cal S})\bigr)$,
from which (D.25a) follows.

\vfill\eject
\reference
\bye